\documentclass[journal,twoside,web]{ieeecolor}
\usepackage{tuffc}
\usepackage{cite}
\usepackage{amsmath,amssymb,amsfonts}
\usepackage{algorithmic}
\usepackage{graphicx}
\usepackage{textcomp}
\usepackage{wrapfig,colortbl}
\usepackage{lipsum}
\usepackage{array}
\usepackage{comment}
\usepackage{threeparttable}

\newcolumntype{M}[1]{>{\centering\arraybackslash}m{#1}}
\usepackage{graphicx}
\usepackage{caption} 
\usepackage{upgreek}

\definecolor{abstractbg}{rgb}{1,0.969,0.914}
\def\BibTeX{{\rm B\kern-.05em{\sc i\kern-.025em b}\kern-.08em
    T\kern-.1667em\lower.7ex\hbox{E}\kern-.125emX}}

\begin{document}

\title{Bimorph Lithium Niobate Piezoelectric Micromachined Ultrasonic Transducers}

\author{Vakhtang Chulukhadze, Zihuan Liu, Ziqian Yao, Lezli Matto, Tzu-Hsuan Hsu, Nishanth Ravi, Xiaoyu Niu, Michael E. Liao, Mark S. Goorsky, Neal Hall, and Ruochen Lu
\thanks{This paper is an expanded version of the IEEE International Ultrasonics Symposium (IUS) 2025. }
\thanks{This work was supported by the Defense Advanced Research Projects Agency (DARPA) High Operational Temperature Sensors (HOTS) project, HR00112420334.}
\thanks{Vakhtang Chulukhadze, Zihuan Liu, Ziqian Yao, Tzu-Hsuan Hsu, Xiaoyu Niu, Neal Hall, and Ruochen Lu are with the University of Texas at Austin, Electrical and Computer Engineering Department.}
\thanks{Lezli Matto, Nishanth Ravi, Michael E. Liao, and Mark Goorsky are with the University of California at Los Angeles, Department of Materials Science and Engineering.}}

\IEEEtitleabstractindextext{%
\fcolorbox{abstractbg}{abstractbg}{%
\begin{minipage}{\textwidth}\rightskip3em\leftskip\rightskip\bigskip
\begin{wrapfigure}[21]{r}{3in}%
\hspace{-5pc}\includegraphics[width=2.9in]{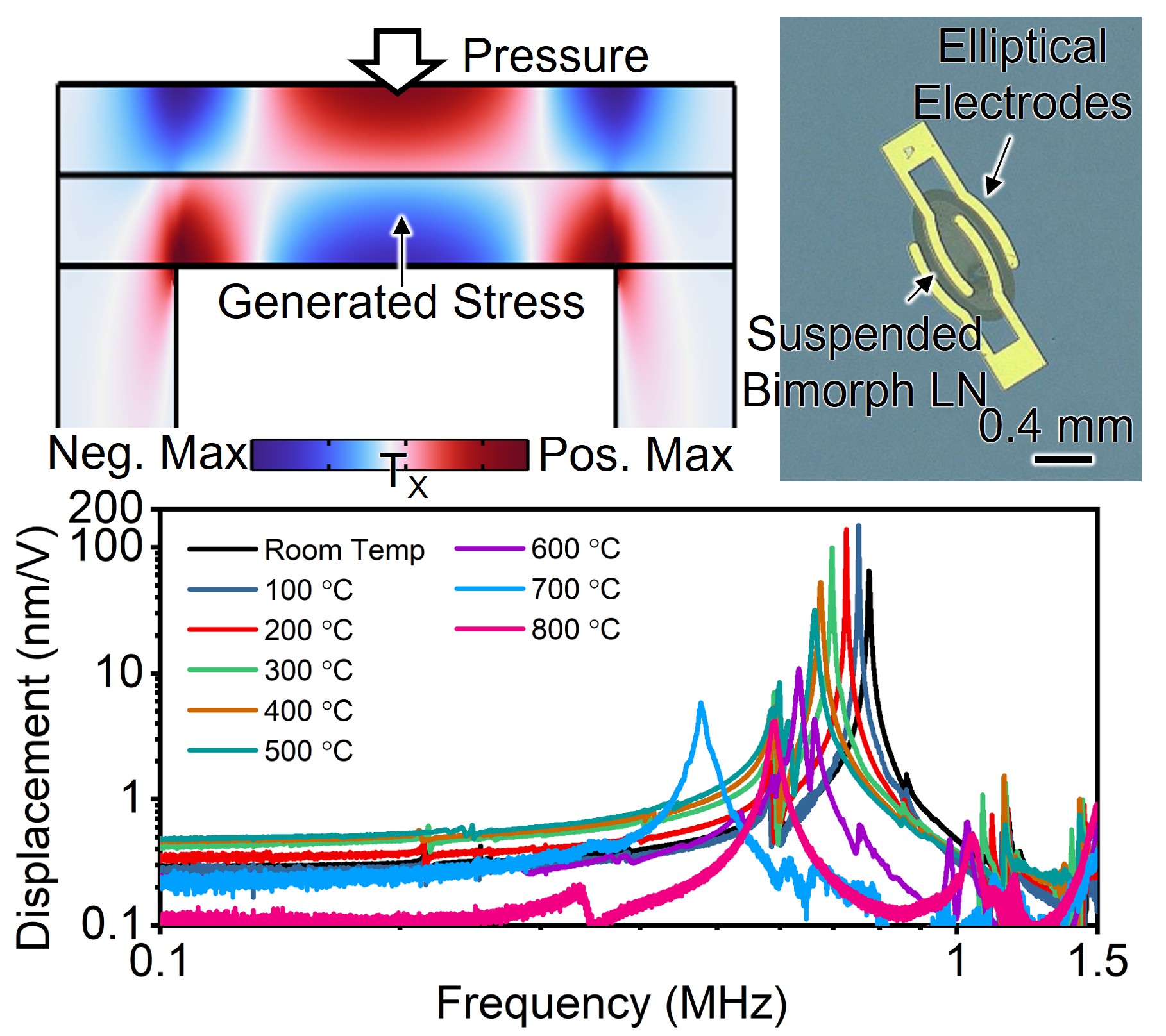}
\end{wrapfigure}%
\begin{abstract}
Piezoelectric micromachined ultrasonic transducers (PMUTs) are widely utilized in applications that demand mechanical resilience, thermal stability, and compact form factors. Recent efforts have sought to demonstrate that single-crystal lithium niobate (LN) is a promising PMUT material platform, offering high electromechanical coupling (\(k^2\)) and bidirectional performance. In addition, advances in LN film transfer technology have enabled high quality periodically poled piezoelectric films (P3F), facilitating a bimorph piezoelectric stack without intermediate electrodes. In this work, we showcase a bimorph PMUT incorporating a mechanically robust, 20 \(\upmu\)m thick P3F LN active layer. We establish the motivation for LN PMUTs through a material comparison, followed by extensive membrane geometry optimization and subsequent enhancement of the PMUT's \(k^2\). We demonstrate a 775 kHz flexural mode device with a quality factor (\(Q\)) of 200 and an extracted \(k^2\) of 6.4\%, yielding a high transmit efficiency of 65 nm/V with a mechanically robust active layer. We leverage the high performance to demonstrate extreme-temperature resilience, showcasing stable device operation up to 600~$^\circ$C and survival up to 900~$^\circ$C, highlighting LN's potential as a resilient PMUT platform.

\end{abstract}

\bigskip
\end{minipage}}}

\maketitle

\begin{table*}[!t]
\arrayrulecolor{subsectioncolor}
\setlength{\arrayrulewidth}{1pt}
{\sffamily\bfseries\begin{tabular}{lp{6.75in}}\hline
\rowcolor{abstractbg}\multicolumn{2}{l}{\color{subsectioncolor}{\itshape
Highlights}{\Huge\strut}}\\
\rowcolor{abstractbg}$\bullet$ & Limitations in incumbent PMUT material platforms motivate an investigation into bimorph LN for ultrasonic transduction.\\
\rowcolor{abstractbg}$\bullet${\large\strut} & This paper demonstrates an LFE PMUT based on a robust 20 $\upmu$m thick transferred P3F bimorph LN on a silicon (Si) carrier wafer. \\
\rowcolor{abstractbg}$\bullet${\large\strut} & A temperature resilience test shows stable PMUT operation and excellent structural stability up to 600~$^\circ$C, alongside survival to 900~$^\circ$C following a sustained extreme temperature bath at 800~$^\circ$C.\\[2em]\hline
\end{tabular}}
\setlength{\arrayrulewidth}{0.4pt}
\arrayrulecolor{black}
\end{table*}

\section{Introduction}
\label{sec:introduction}
\IEEEPARstart{P}{iezoelectric} micromachined ultrasonic transducers (PMUTs) are widely used in applications that require mechanical durability, thermal resilience, and compact form factors. Examples include range-finding, biomedical imaging, high-temperature sensors,  and fingerprint sensors \cite{luo_airborne_2021, chen_high-accuracy_2019, hunt_ultrasound_1983, zhang_piezoelectric_2011, jiang_ultrasonic_2017}. Typical PMUT materials include lead zirconate titanate (PZT), potassium sodium niobate (KNN), barium titanate (BaTiO\(_\text{3}\)), zinc oxide (ZnO), aluminum nitride (AlN), and scandium aluminum nitride (ScAlN), each offering different trade-offs in actuation, sensing, bidirectional performance, and environmental resilience \cite{belgacem_piezoelectric_2006, xia_high_2024,acosta_batio3-based_2017, baumgartel_use_2010, williams_aln_2012, wang_design_2017}. 

\begin{figure} [t!]
\centering
\includegraphics[width=\columnwidth]{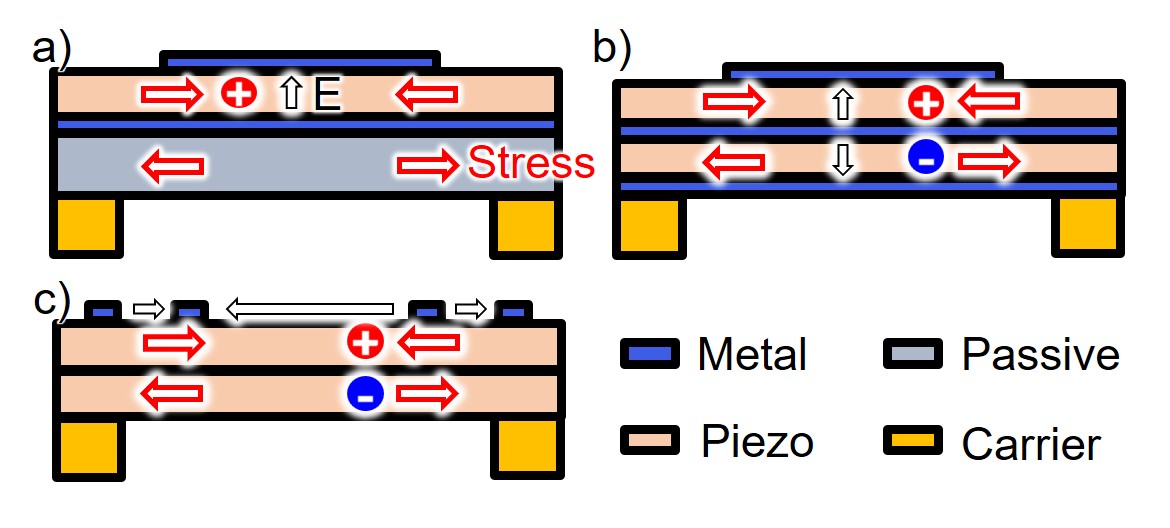}
\caption{Cross-sectional views of a conventional thickness field excited PMUT in (a) unimorph configuration with a passive layer. (b) Bimorph configuration with three electrode layers. (c) Proposed P3F bimorph configuration excited with lateral electrical fields using a single electrode layer.}
\label{UnimorphBimorph}
\end{figure}

PMUTs conventionally rely on bilayer (bimorph) active layers to maximize transduction efficiency \cite{akhbari_bimorph_2016}. This configuration aims to compensate for the intrinsic stress distribution during flexural motion, thereby preventing full charge cancellation in a single-layer (unimorph) piezoelectric flexural mode device. While an alternative unimorph structure can employ passive mechanical layers to mitigate this effect, it introduces substantial performance bottlenecks: a reduced electromechanical coupling (\(k^2\)) due to stress in the passive layer, and increased thermal and mechanical stress resulting from its mismatched stiffness and thermal expansion coefficients with the active layer. These configurations are summarized in Fig. \ref{UnimorphBimorph} (a) and (b). Consequently, beyond intrinsic material constants, the ability to fabricate high-quality bimorph acoustic stacks is a critical determinant of PMUT performance.   

PZT, a lead-based perovskite, remains dominant in the industry owing to its strong transmit characteristics \cite{tsang_piezoelectric_2002}. However, its high electrical permittivity at constant stress (\(\varepsilon^T\)) and the high associated dielectric loss limit sensing performance, such as the signal-to-noise ratio \cite{tressler_piezoelectric_1998}. Despite these drawbacks, the maturity of PZT fabrication processes continues to sustain its role as a primary PMUT material. To overcome the limitations of PZT, AlN is employed for its low dielectric loss and high sensitivity. Although it possesses moderate piezoelectricity, leading to moderate \(k^2\), its low \(\varepsilon^T\) and low dielectric loss enhance sensitivity and signal-to-noise ratio, making it ideal for low-power, high-sensitivity systems \cite{shao_bimorph_2021, akhbari_highly_2014}. Nevertheless, PZT and AlN exhibit complementary limitations: PZT excels as an actuator but performs poorly as a sensor, whereas AlN offers high sensitivity but limited actuation. A comparison is listed in Tables \ref{tab1} and \ref{tab2} \cite{qiao_full_2020, shi_enhanced_2020, caro_erratum_2015, wolff_al1xscxn_2022, streque_stoichiometric_2019}.

Recent advancements in piezoelectric materials aim to overcome the inherent trade-offs among transmission efficiency, sensitivity, and environmental compliance by pursuing lead-free alternatives. Potassium sodium niobate (KNN), a lead-free perovskite, offers improved actuation, but despite efforts to minimize \(\varepsilon^T\), KNN remains sub-optimal for an acoustic sensor \cite{xia_high_2024}. By contrast, ScAlN has been developed to enhance AlN's piezoelectric response through scandium alloy incorporation, significantly increasing its piezoelectric coefficients at the cost of slightly higher \(\varepsilon^T\) and dielectric loss. Regardless, ScAlN remains less piezoelectric than PZT, KNN, and other incumbent platforms. Moreover, both KNN and ScAlN are constrained by thin-film deposition processes, which limit the achievable film thickness and complicate bimorph stack formation without incorporating intermediate electrode layers \cite{iborra_piezoelectric_2004, lundh_residual_2021, shibata_knn_2022, xia_high-spl_2023}. These challenges underscore the need for an alternative piezoelectric platform that can strike a balance in bidirectional acoustic performance.

\begin{table} 
\centering
\caption{Bulk PMUT Material Parameters}
\label{tab1}
\begin{threeparttable}
\renewcommand{\arraystretch}{1.2}
\setlength{\tabcolsep}{4pt}
\begin{tabular}{
  M{45pt}
  M{25pt}
  M{25pt}
  M{25pt}
  M{25pt}
  M{25pt}
  M{25pt}
}
\hline\hline
Material & Excita-tion & \(d_\textit{ij}\) (pm/V) & \(g_\textit{ij}\) (V/m/Pa) & \(e_\textit{ij}\) (C/m\(^2\)) & \(\varepsilon_{r_\textit{ii}}^{T}\) & \(T_C\) \\
\hline
PZT-5A & TFE & -171 & -0.011 & -5.4 &  1700 & 350~$^\circ$C\\
KNN & TFE & -140 & -0.008 & -11.2 & 2000 & 235~$^\circ$C\\
Sc\(_\text{36}\)Al\(_\text{64}\)N & TFE & 8.9 & 0.030 & -0.7 & 33.64 & 1100~$^\circ$C \\
36$^\circ$Y LN & TFE & -16.9 & -0.024 & -1.8 & 65.2 & 1200~$^\circ$C\\
36$^\circ$Y LN & LFE & 38.5 & 0.054 & 4.6 & 67.2 & 1200~$^\circ$C\\
128$^\circ$Y LN & TFE & -27.0 & -0.033 & -2.7 &  50.5 & 1200~$^\circ$C\\
128$^\circ$Y LN & LFE & -37.4 & -0.053 & -4.5 & 63.5 & 1200~$^\circ$C\\
Y-cut LN & TFE & -20.3 & -0.027 & -2.4 & 84.0 & 1200~$^\circ$C\\
Y-cut LN & LFE & 27.1 & 0.040 & 3.2 & 58.3 & 1200~$^\circ$C\\
\textbf{X-cut LN} & LFE & -40.0 & -0.054 & -4.6 & 69.7 & 1200~$^\circ$C \\
\hline\hline
\end{tabular}
\begin{tablenotes}[flushleft]
\footnotesize
\item *The subscripts are \(31\) for TFE excitation  and \(11\) for LFE excitation.
\end{tablenotes}
\end{threeparttable}
\end{table}

\begin{table} 
\centering
\caption{Thin-film PMUT Material Parameters}
\label{tab2}
\begin{threeparttable}
\renewcommand{\arraystretch}{1.2}
\setlength{\tabcolsep}{4pt}
\begin{tabular}{
  M{85pt}
  M{33pt}
  M{33pt}
  M{33pt}
  M{33pt}
}
\hline\hline
Material & Excitation & $e_{ij\_\text{f}}$ (C/m$^2$) & $\varepsilon_{r_{ii}\_\text{f}}$ & \(e_{ij\_\text{f}}\)/\(\varepsilon_{{ii}\_\text{f}}\) (GV/m) \\
\hline
Epitaxial PZT & TFE & -24.0 & 308 & -8.8  \\
Sol-gel PZT & TFE & -17.7 & 1076 & -1.9 \\
Sputtered KNN & TFE & -10.0 & 260 & -4.3 \\
Sputtered Sc\(_\text{36}\)Al\(_\text{64}\)N & TFE & -2.4 & 24.4 & -12.5 \\
\textbf{Transferred X-cut LN} & LFE & -5.3 & 41.1 & -14.6\\
\hline\hline
\end{tabular}
\begin{tablenotes}[flushleft]
\footnotesize
\item *The subscripts are \(31\) for TFE excitation  and \(11\) for LFE excitation.
\end{tablenotes}
\end{threeparttable}
\end{table}

Single-crystal piezoelectric lithium niobate (LN) has recently emerged as a promising candidate to address these limitations~\cite{lu_piezoelectric_2020, niu_lithium_2025}. LN offers a balanced combination of high piezoelectric coupling coefficients and a moderate \(\varepsilon^T\), situating it between KNN and ScAlN \cite{lu_recent_2025, chulukhadze_planar_2025, chulukhadze_cross-sectional_2025, chulukhadze_high-q_2025,chulukhadze_toward_2026}. Lu \(\textit{et al.}\) first demonstrated a unimorph lateral-field-excited (LFE) acoustic transducer with a silicon oxide (SiO\(_2\)) passive layer, achieving a high \(k^2\) of 4.5\% and illustrating LN's potential for efficient acoustic transduction \cite{lu_piezoelectric_2020}. Importantly, unlike KNN and ScAlN, LN is primarily available as a bulk single-crystal, allowing for precise thickness control via smart cut or controlled chemical-mechanical polishing (CMP) \cite{jia_ion-cut_2021, wu_thinning_2006}. Moreover, recent advances in LN-LN bonding technology now enable the production of high-quality bimorphs using periodically poled piezoelectric layers (P3F) [Fig. \ref{UnimorphBimorph} (c)], which do not necessitate inter-layer electrodes, thereby expanding design flexibility and fabrication reliability \cite{barrera_fundamental_2023, kramer_thin-film_2023, zheng_periodically_2025}. These developments motivate a revisit of the LN PMUT.

P3F LN differs from conventional piezoelectric platforms in its ability to efficiently excite the flexural mode without the need for additional electrodes between active layers. Due to its anisotropy, the LN wafer orientation can be flipped during the LN-LN wafer bonding process to change the polarity of the appropriate piezoelectric coupling constant, effectively forming a bimorph stack without the need for electrical fields of opposite polarity. A theoretical framework regarding multi-layer P3F structures in anisotropic materials can be found in \cite{kramer_generalized_2025}. While P3F LN has been predominantly used in radio frequency (RF) acoustic applications, efforts to leverage its benefits for ultrasonic transduction have achieved significant success \cite{barrera_50_2026, cho_238-ghz_2024,niu_lithium_2025, zhao_piezoelectric_2025, zhao_optimal_2025}. We aim to further develop this platform by providing a comprehensive theoretical and experimental framework around LN PMUT design and its capabilities.

This manuscript expands on work presented at the 2025 IEEE International Ultrasonics Symposium (IUS)~\cite{yao_bimorph_2025}. In comparison, this study presents a comprehensive theoretical, design, and experimental investigation of a bimorph LN PMUT. The study begins by evaluating the figures of merit (FoM) of incumbent PMUT platforms, establishing the theoretical motivation for LN-based devices. Consequently, a design and simulation framework is developed that accounts for LN's in-plane anisotropy. The device is fabricated using a 20\(~\mu\text{m}\) thick active piezoelectric layer on silicon (Si) with a silicon oxide (SiO\(_2\)) interlayer. The experimental results demonstrate high transmit efficiency and show stable operation up to 600~$^\circ$C, with survival to approximately 900~$^\circ$C. Thus, we utilize the bimorph LN platform to demonstrate high-performance PMUTs capable of sustained operation under harsh environmental conditions. 

\section{Design and Simulation}

\begin{figure} [t!]
\centering
\includegraphics[width=\columnwidth]{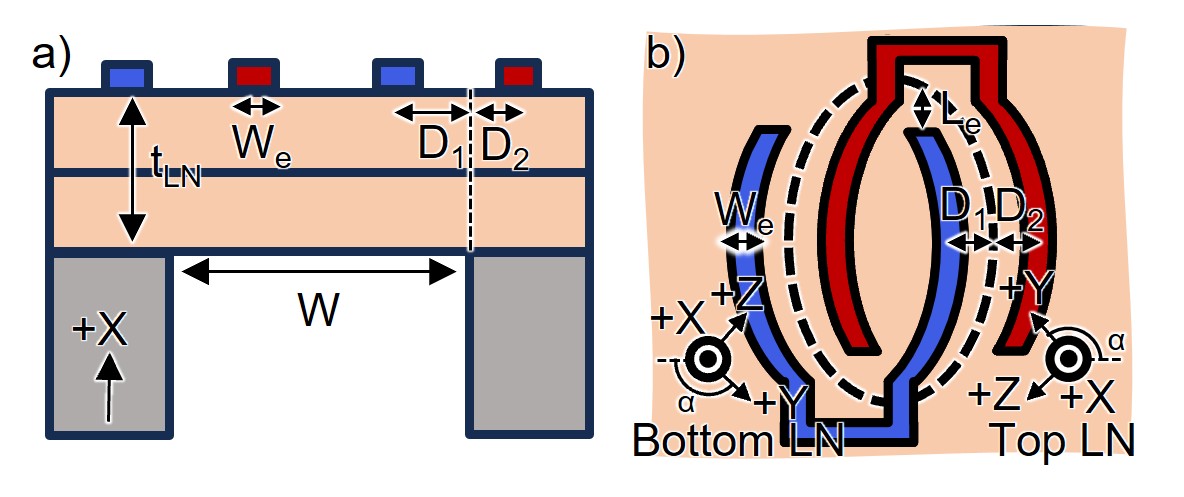}
\caption{(a) Cross-sectional schematic of a basic LFE LN PMUT and its design parameters. (b) Top-view schematic of the proposed LN PMUT design, highlighting the LFE excitation mechanism and an elliptical membrane shape.}
\label{3Dschem}
\end{figure}

\begin{table}[!t]
\caption{Chosen Key Design Parameters}
\label{tab:geom_params}
\centering
\renewcommand{\arraystretch}{1.2}
\setlength{\tabcolsep}{4pt}
\small
\begin{tabular*}{\columnwidth}{@{\extracolsep{\fill}}ccccccc}
\hline
$t_{LN}$& $W$ & $L$ & $D_1$ & $D_2$ & $W_e$ & $L_e$ \\
\hline
20\(~\mu\text{m}\) & 360\(~\mu\text{m}\) & 760\(~\mu\text{m}\) & 45\(~\mu\text{m}\) & 21\(~\mu\text{m}\) & 60\(~\mu\text{m}\) & 200\(~\mu\text{m}\) \\
\hline
\end{tabular*}
\end{table}

We aim to leverage P3F LN by employing a 20 $\upmu$m thick bimorph X-cut LN on 400 nm silicon oxide (SiO\(_2\)) on 200 $\upmu$m silicon (Si) stack. The chosen film thickness balanced mechanical robustness and compactness while exceeding the limitations of conventional thin-film deposition techniques used for PZT, KNN, or ScAlN. The cross-sectional and top-views of the chosen acoustic stack can be seen in Fig. \ref{3Dschem} (a) and (b), illustrating the bimorph active layer achieved via P3F LN, and the electrical excitation scheme adopted from Lu \(\textit{et al.}\)~\cite{lu_piezoelectric_2020}. 

\begin{figure}[!t]
\centerline{\includegraphics[width=\columnwidth]{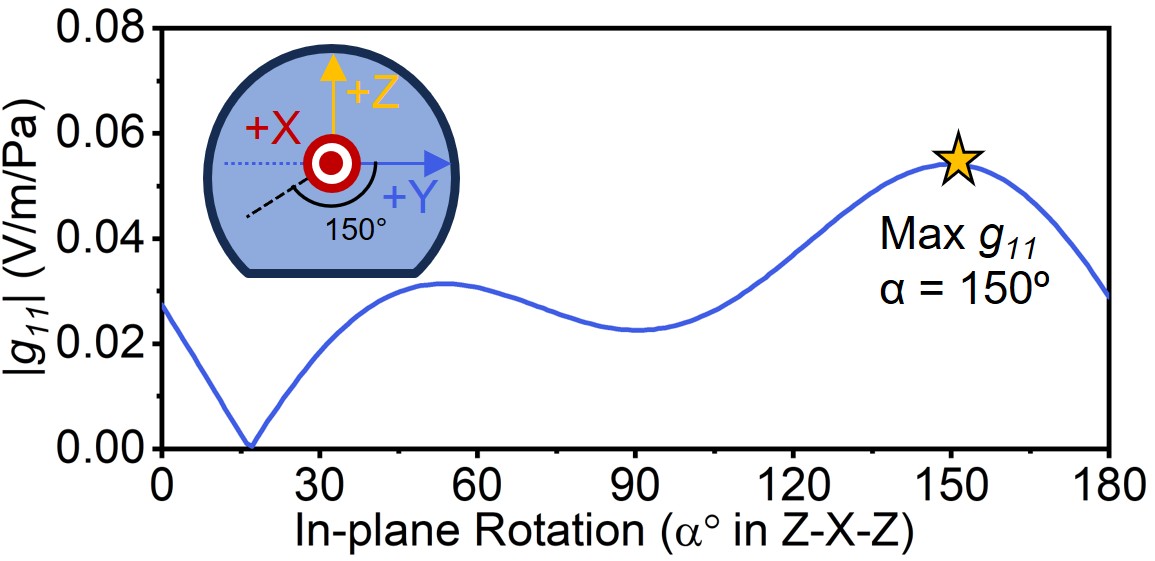}}
\caption{\(g_\textit{11}\) electromechanical coupling constant in X-cut LN against wafer in-plane rotation (\(\alpha\)).}
\label{Orientation}
\end{figure}

\begin{figure}[t!]
\centering
\includegraphics[width=\columnwidth]{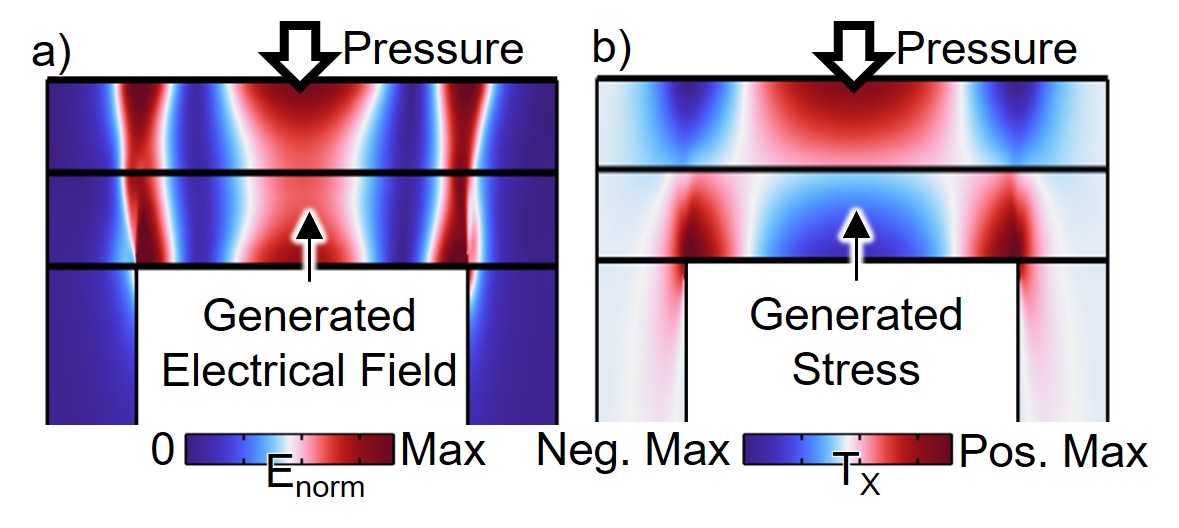}
\caption{Cross-sectional views of (a) the generated electrical field in a P3F LN membrane due to a unit pressure input. (b) The longitudinal stress induced by a pressure load in the same geometry, showing that the electrical field reflects the stress profile.}
\label{Crosssection_Enorm_Stress}
\end{figure}

\begin{figure}[t!]
\centering
\includegraphics[width=\columnwidth]{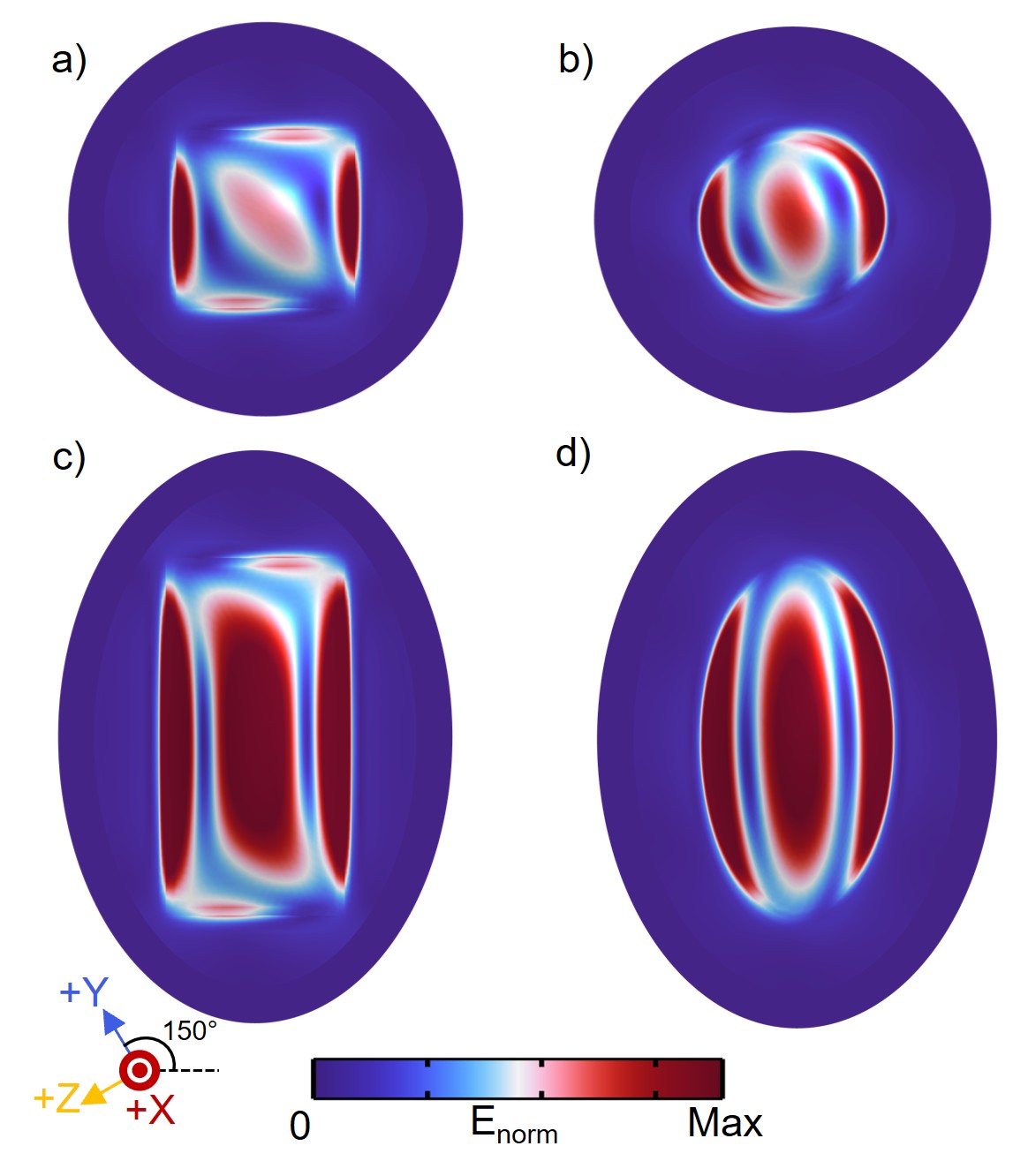}
\caption{Top-view perspective of the excited open-circuit voltage due to a pressure boundary load in a (a) square, (b) circular, (c) rectangular, and (d) elliptical membranes, showcasing how (b) and (d) reduce the excitation of the parasitic transverse fields.}
\label{Shape}
\end{figure}

The design process began with a comparison of PZT-5A, KNN, 36\% ScAlN, and different LN cuts, which motivated our approach. The LN material orientation was optimized for a fair comparison (Fig.~\ref{Orientation}), and utilized in COMSOL finite element analysis (FEA) to define design parameters while accounting for in-plane anisotropy in LN (Figs.~\ref{Crosssection_Enorm_Stress} and \ref{Shape}). Finally, the obtained parameters were used to predict the acoustic transducer's transmit efficiency. 

To justify the selection of X-cut LN for acoustic transduction, material FoMs were compared across incumbent piezoelectric platforms (Table~\ref{tab1}). The primary metrics evaluated the material's piezoelectric response under different electrical and mechanical boundary conditions. The strain-charge piezoelectric coupling tensor (\(d\), in m/V), and the strain-voltage piezoelectric coupling tensor (\(g\), defined as \(d/\varepsilon^T\) in V/m/Pa) were used to quantify mechanical strain generation due to an electrical input, and electrical field generation due to a mechanical input, respectively  \cite{jaffe_piezoelectric_1965}. While these tensors describe the material when it is free to deform, the stress-charge piezoelectric coupling tensor (\(e\), in C/m\(^2\)) relates the electrical field with stress when the material is clamped. Accordingly, \(d\) and \(e\) are widely adopted to evaluate material's transmit efficiency, depending on physical boundary conditions. On the other hand, characterizing receiving sensitivity depends on the measurement configuration. Often, \(e\) is preferred over \(g\) for materials with large capacitance densities, making charge read-out the optimum interfacing method with the device. As a result, \(d\), \(g\), and \(e\) were utilized as FoMs to characterize the material's piezoelectric response. The relative \(\varepsilon^T\) (\(\varepsilon_r^T\)) and Curie temperature (\(T_C\)) served as secondary metrics to evaluate capacitance density and robustness.

The comparison in Table~\ref{tab1} highlights key performance trade-offs among the bulk material constants of PZT-5A, KNN, 36\% ScAlN, and LN, motivating our choice of piezoelectric material. PZT-5A demonstrates the highest \(d\) coupling tensor coefficient, but possesses a large \(\varepsilon^T\) and a low \(T_C\), limiting sensing performance and resilience. KNN features a comparable \(d\) and the highest \(e\) coefficients, but remains sub-optimal compared with ScAlN and LN due to its high \(\varepsilon^T\). ScAlN achieves high \(g\) and \(T_C\), but it falls short of PZT and KNN in transmit FoMs. In contrast, different cuts of LN exhibit a very high \(g\), paired with a moderate \(d\), and a high \(T_C\), which motivates its use for robust, high-temperature PMUTs. Moreover, LN offers unprecedented design flexibility, supporting both thickness-field-excited (TFE) and LFE electrode configurations.

The bulk piezoelectric constants seen in Table~\ref{tab1} are significantly affected when a carrier substrate is introduced \cite{dubois_measurement_1999}. In this configuration, the material is clamped in-plane but free to deform out-of-plane, allowing additional tensor components to contribute to the measured piezoelectric response. Namely, an effective stress-charge piezoelectric coupling constant (\(e_{ij\_\text{f}}\), in C/m\(^2\)) and an effective permittivity (\(\varepsilon_{{ii}\_\text{f}}\)) can be defined. Their ratio (\(e_{ij\_\text{f}}\)/\(\varepsilon_{{ii}\_\text{f}}\), in GV/m) describes the generated voltage in the constrained film, acting as a receive FoM. Several material platforms, such as PZT, KNN, and ScAlN, exhibit substantial enhancement of their effective piezoelectric coefficients under in-plane clamping, as summarized in Table \ref{tab2} \cite{luo_airborne_2021, calame_growth_2007, xia_high_2024, mertin_enhanced_2017}. For example, the reported  \(e_\text{\textit{31}\_\textit{f}}\) of -2.4 C/m\(^2\) in 36\% ScAlN is markedly larger than its pure \(e_\text{\textit{31}}\) of only -0.7 C/m\(^2\), which was experimentally verified in \cite{mertin_enhanced_2017}. To compare with these measurements, the effective material constants for LFE PMUTs in X-cut LN were derived from the stress-charge piezoelectric constitutive system while assuming zero in-plane strain and zero out-of-plane stress. The resulting parameters, shown in Table \ref{tab2}, indicate that X-cut LN maintains an advantage over incumbent platforms under substrate clamping, reinforcing the suitability of P3F X-cut LN for bi-directional PMUT operation in both bulk and thin-film platforms.

\begin{table}[!t]
\caption{\(g\) piezoelectric coupling tensor in X-cut LN with a 150$^\circ$ in-plane rotation.}
\label{tab3}
\centering
\setlength{\tabcolsep}{4pt}
\renewcommand{\arraystretch}{1.15}
\begingroup
{\fontsize{12}{14}\selectfont }$%
g=\left[
\begin{array}{rrrrrr}
-0.054 &  0.021 &  0.025 &  0      & 0      & -0.028 \\
 0.028 & -0.031 & -0.011 &  0      & 0      & -0.010 \\
 0     &  0     &  0     & -0.108  & 0      & 0
\end{array}
\right]\;\frac{\mathrm{V}}{\mathrm{m\cdot Pa}}%
$
\endgroup
\end{table}

LN is available in a variety of different material cuts, each with a unique set of electromechanical coupling constants and electrical characteristics. Table \ref{tab1} shows varying excitation mechanisms and FoMs between 36$^\circ$Y, 128$^\circ$Y, Y-cut, and X-cut LN platforms. However, although the table shows similar \(T_C\) for all LN cuts, high-temperature stability investigations of LN have shown a premature phase segregation point at 800~$^\circ$C in congruent LN \cite{streque_stoichiometric_2019}. Meanwhile, LN with a stoichiometric composition shows no such detriments until \(T_C\) is reached. Importantly, only orthogonal LN cuts are commercially available with a stoichiometric composition. Hence, X-cut LN is the primary choice for an LFE PMUT platform due to its high FoMs and potential for high temperature resilience.  

To design piezoelectric transducers using X-cut LN, we first examined the \(g_\textit{11}\) coupling constant as a function of wafer in-plane orientation. Among the bottom and top layer orientations seen in Fig.~\ref{3Dschem} (b), we utilized the top LN layer to describe the device rotation. As a result, we found that an electrical field oriented 150\(^\circ\) from the crystal +Y axis maximizes \(g_\textit{11}\) (Fig.~\ref{Orientation}). Due to the P3F structure, the same results, with an additional 180$^\circ$ in-plane rotation, apply to the bottom LN layer.

The pronounced in-plane anisotropy in LN necessitates reconsideration of diaphragm shape and geometry. An FEA was used to compare square, circular, rectangular, and elliptical shapes under uniform pressure loading, shown in Fig.~\ref{Crosssection_Enorm_Stress} and Fig.~\ref{Shape}, for the cross-sectional and top views, respectively. A 180$^\circ$ in-plane rotation between the top and bottom LN layers yields the P3F structure. The corresponding spatial distribution of the generated electrical field draws important insights in context with the full \(g\) piezoelectric coupling tensor in X-cut LN (Table \ref{tab3}). Fig. \ref{Shape} (a) and (c) demonstrate how a square or a rectangular diaphragm generates a complex electrical field with both longitudinal and transverse components. However, due to the opposing signs of \(g_\textit{11}\) and \(g_\textit{12}\), the transverse field is parasitic, leading to partial cancellation with the desired longitudinal field. In contrast, Fig. \ref{Shape} (b) and (d) show that a circular shape mitigates parasitic effects, while an elliptical diaphragm further suppresses transverse anti-nodes. Thus, an elliptical geometry was adopted for subsequent optimization. 


\begin{figure}[!t]
\centerline{\includegraphics[width=\columnwidth]{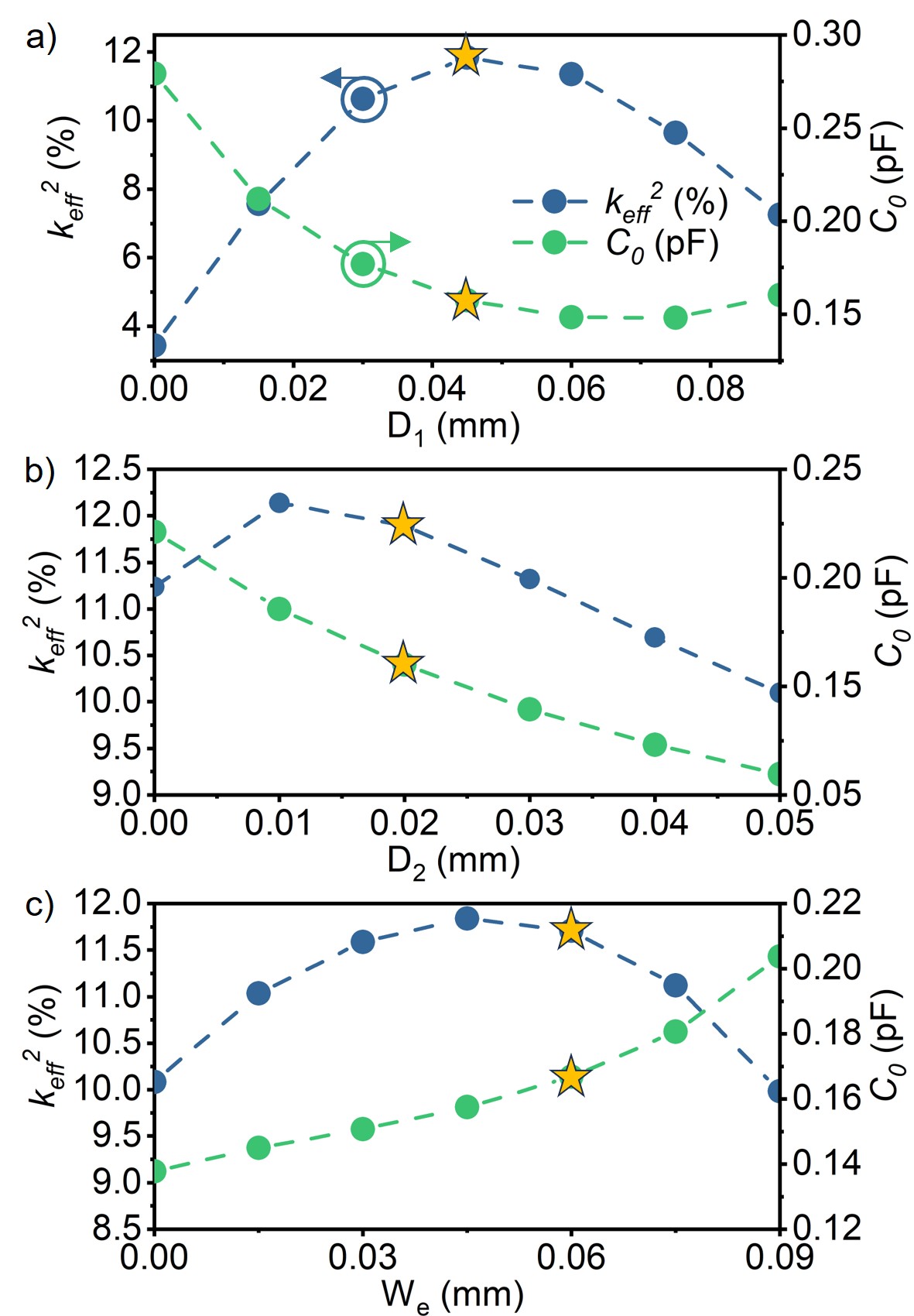}}
\caption{(a) \(k^2_\textit{eff}\) and \(C_0\) as a function of electrode distance \(D_1\). (b) \(k^2_\textit{eff}\) and \(C_0\) as a function of electrode distance \(D_2\). (c) Electrode width (\(W_e\)) against \(k^2_\textit{eff}\) and \(C_0\). Besides the parameter actively being swept, all other key design parameters are held constant at their values from Table \ref{tab:geom_params}. The chosen parameter values are labeled with a star.}
\label{k2C0}
\end{figure}

The chosen diaphragm geometry introduces new key design parameters in addition to those reported by Lu \(\textit{et al.}\), shown in Fig. \ref{3Dschem} (b) \cite{lu_piezoelectric_2020}. Namely \(L_e\), and \(A\). \(A\) is defined as the ratio of the ellipse's minor axis (\(W\)), to its major axis (\(L\)), i.e., \(A\) = \(W/L\). The parameter \(L_e\) defines the transverse electrode offset, optimized to match the transverse energy distribution in the diaphragm while also providing space for routing to contact pads.

\begin{figure}[!t]
\centerline{\includegraphics[width=\columnwidth]{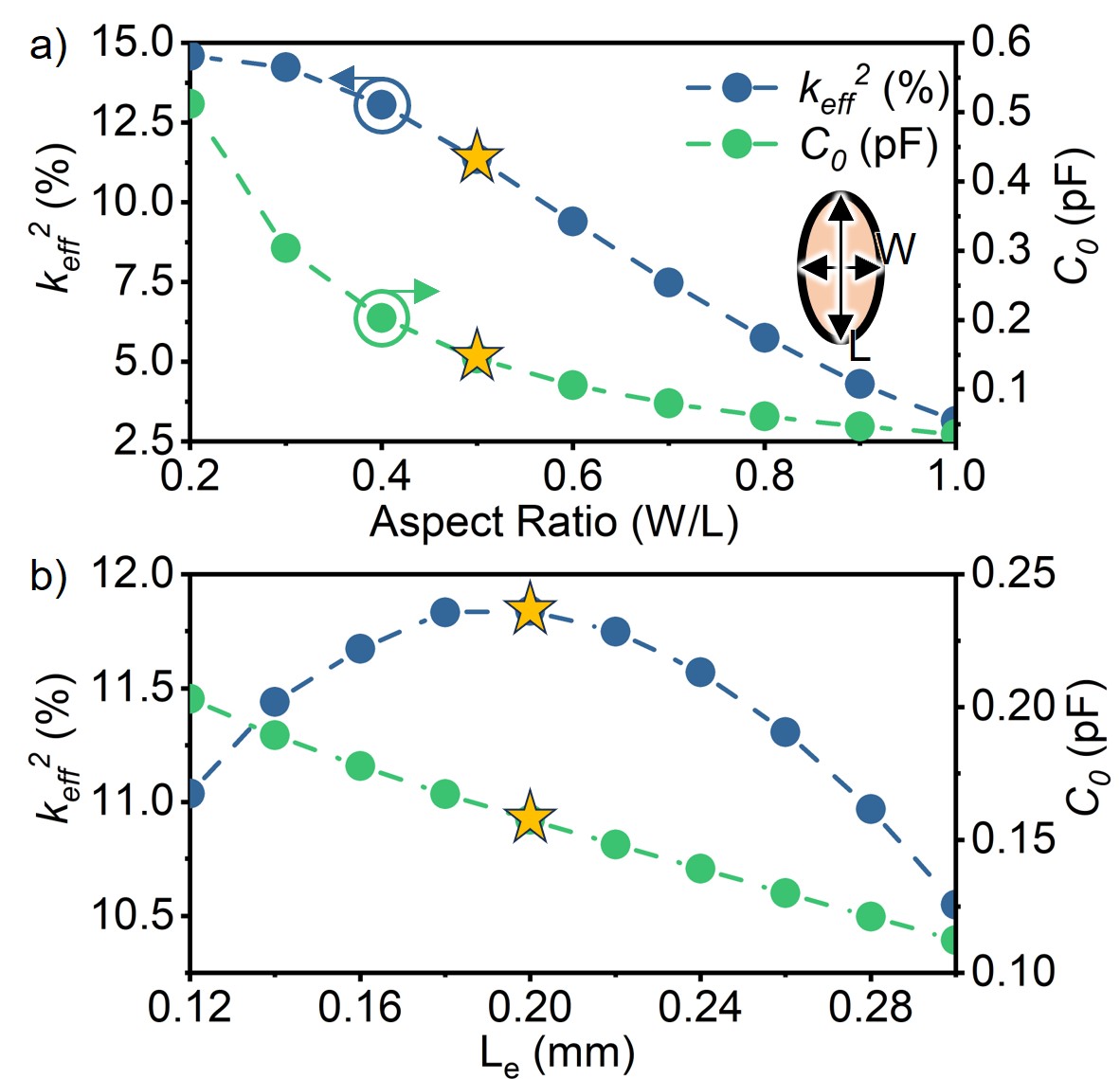}}
\caption{(a) \(k^2_\textit{eff}\) and \(C_0\) as a function of the elliptical membrane aspect ratio (\(A\)), indicating how a small aspect ratio leads to a higher \(k^2_\textit{eff}\) and \(C_0\). (b) \(k^2\) and \(C_0\) as a function of the transverse electrode offset (\(L_e\)). Besides the parameter actively being swept, all other key design parameters are held constant at their values from Table \ref{tab:geom_params}. The chosen parameter values are labeled with a star.}
\label{k2A}
\end{figure} 

\begin{figure}[!t]
\centerline{\includegraphics[width=\columnwidth]{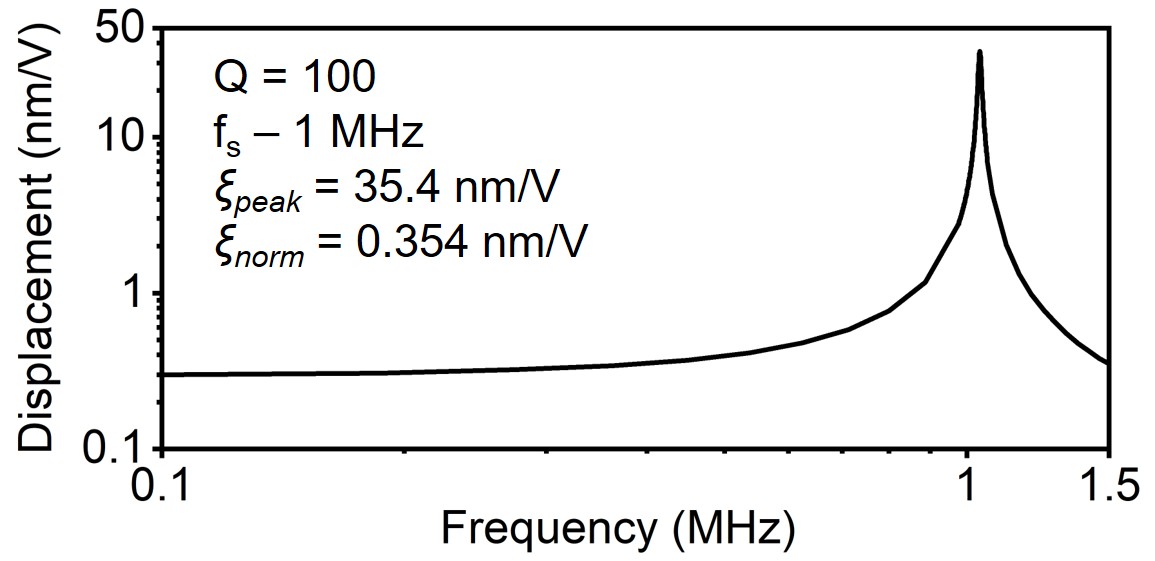}}
\caption{Simulated peak displacement for a \(Q\) of 100 given the design specifications from Table \ref{tab:geom_params}.}
\label{SimDisp}
\end{figure} 

A COMSOL FEA was used to obtain the key design parameters for optimized operation at 1 MHz, listed in Table \ref{tab:geom_params}. \(k^2_\textit{eff}\) and static capacitance \(C_0\) were tracked as the primary performance metrics. \(k^2_\textit{eff}\) determines electromechanical transduction efficiency for a resonant structure, rather than the material itself \cite{berlincourt_piezoelectric_1964}. For a PMUT with voltage and pressure as the primary effort variables, it can be found as follows:
\begin{equation}
k^2_\textit{eff} = \Phi^2\cdot C_\textit{ad}\cdot C_0,
\label{eq:k2}
\end{equation}
where $\Phi$ corresponds to the turns ratio of a pressure-voltage electromechanical transformer, \(C_\textit{ad}\) stands for acoustic compliance, and \(C_0\) describes static capacitance. Due to the moderate \(\varepsilon^T\) of LN and its large chosen thickness, voltage sensitivity is preferred~\cite{hindrichsen_advantages_2010}. Hence, $\Phi$ here can be represented as open-circuit voltage normalized by volumetric displacement. Alongside \(k^2_\textit{eff}\), \(C_0\) was tracked as a secondary performance metric due to its critical influence on the transducer's electrical impedance and system-level integration \cite{littrell_modeling_2012}.

The operating frequency was chosen to demonstrate a versatile design framework rather than targeting a specific application. The provided design parameters can be linearly scaled to achieve a resonance frequency in the lower kHz regime to leverage low attenuation, or higher MHz frequencies for superior axial resolution \cite{peng_9-meter-long_2024, koh_high_2021}.

Parameter sweeps (Figs. \ref{k2C0} and \ref{k2A}) isolate the influence of each design variable while holding all others constant (Table \ref{tab:geom_params}). \(D_1\) was tuned for maximum \(k^2_\textit{eff}\) while \(D_2\) and \(W_e\) were limited by fabrication alignment tolerances. To design the aspect ratio without drastic changes to the operating frequency, \(L\) was swept while \(W\) was held constant. Fig. \ref{k2A} (a) shows that a smaller aspect ratio continuously improves both \(k^2_\textit{eff}\) and \(C_0\); however, etching such patterns is challenging. Hence, we nominally selected \(A\) = 0.5 as a compromise between performance and manufacturability. \(L_e\) was selected to co-optimize \(k^2_\textit{eff}\) and \(C_0\) by following the transverse electrical field distribution seen in Fig. \ref{Shape} (d) while also allowing to trace out measurement pads. 

The optimized design parameters in Table \ref{tab:geom_params} were utilized to simulate the transmission efficiency of the resulting device (Fig. \ref{SimDisp}). The simulation in Fig. \ref{SimDisp} predicts a peak transmit efficiency (\(\xi_\textit{peak}\)) of 35.4~nm/V at a nominally selected \(Q\) of 100. To decouple \(Q\) from transmit efficiency, we utilized the peak displacement divided by \(Q\) as the normalized displacement (\(\xi_\textit{norm}\)), yielding a value of 0.354 nm/V. These results confirmed efficient transduction with a thick bimorph LN PMUT, validating the effectiveness of the device design process. Accordingly, the validated design parameters from Table \ref{tab:geom_params} were used to proceed with the device fabrication.

\section{Fabrication and Measurement}
\subsection{Device Fabrication}

\begin{figure}[!t]
\centerline{\includegraphics[width=\columnwidth]{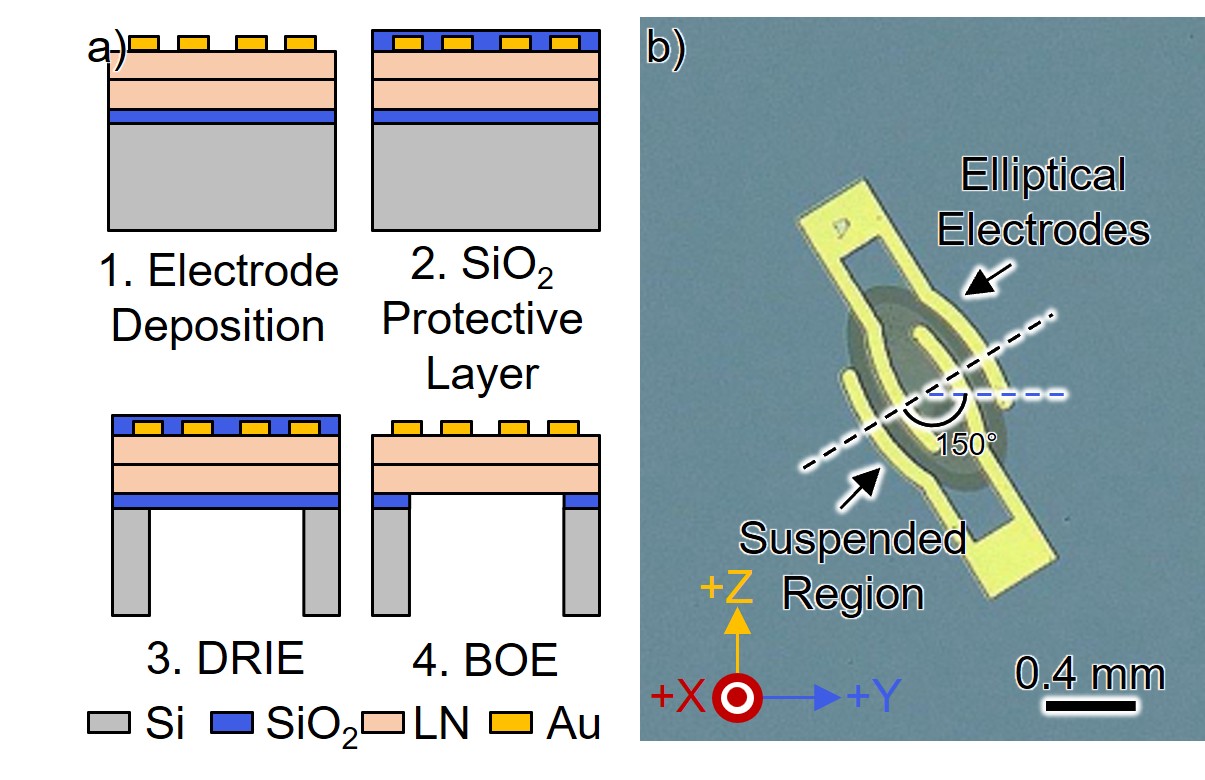}}
\caption{(a) The fabrication flow for this work. (b) An optical image of the resulting bimorph LFE PMUT.}
\label{fab}
\end{figure}

\begin{figure}[!t]
\centerline{\includegraphics[width=\columnwidth]{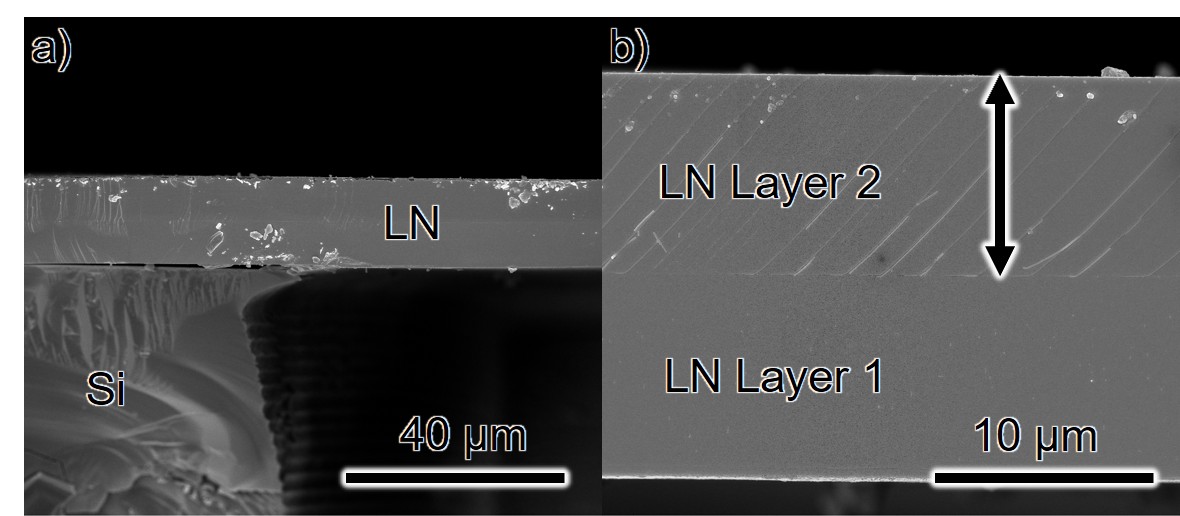}}
\caption{Cross-sectional SEM images of the controlled die depicting (a) the suspended membrane over the Si cavity, and (b) magnified view of the bimorph active layer.}
\label{SEM}
\end{figure}

\begin{figure}[!t]
\centerline{\includegraphics[width=\columnwidth]{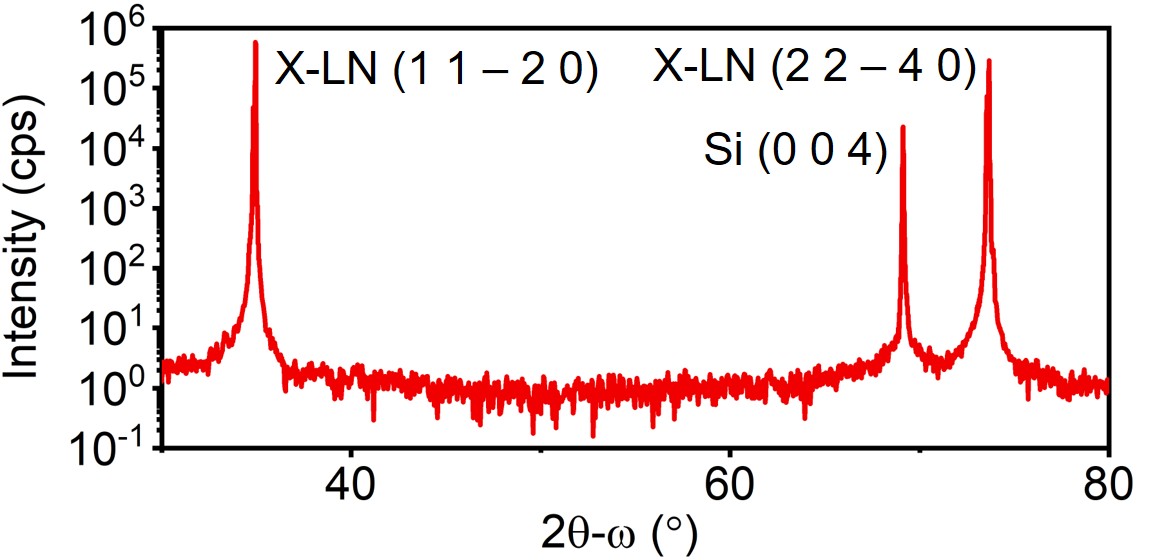}}
\caption{A double-axis 2$\theta$-$\omega$ XRD scan of the acoustic stack, confirming the single phase of the piezoelectric layer.}
\label{XRD_2theta}
\end{figure}

The device was fabricated using a 20 $\upmu$m thick bimorph X-cut LN on 400 nm SiO\(_2\) on a 200 $\upmu$m Si stack from NGK Insulators Ltd. Fabrication occurred with a 3$\times$3 mm\(^\text{2}\) die. Top electrodes were defined by standard photolithography and electron-beam evaporation of a Ti/Pt/Au (10/30/60 nm) stack. The metal stack-up was chosen due to its high melting temperature. A 100 nm electrode thickness was selected to reduce electrical resistance and mass-loading effects. The membrane was released using an Oxford H300 deep reactive ion etching (DRIE) tool, leveraging a (SiO\(_2\)) protective layer deposited using plasma-enhanced chemical vapor deposition (PECVD) to mount the sample. The etch leveraged the 400 nm  (SiO\(_2\)) layer as the etch stop. Following DRIE, the sample was etched using a buffered oxide etch (BOE) to remove remaining SiO\(_\text{2}\). The fabrication flow can be seen in Fig. \ref{fab} (a), and the resulting optical image of the device can be seen in Fig. \ref{fab} (b). 

\subsection{Material-Level Analysis}

\begin{figure}[!t]
\centerline{\includegraphics[width=\columnwidth]{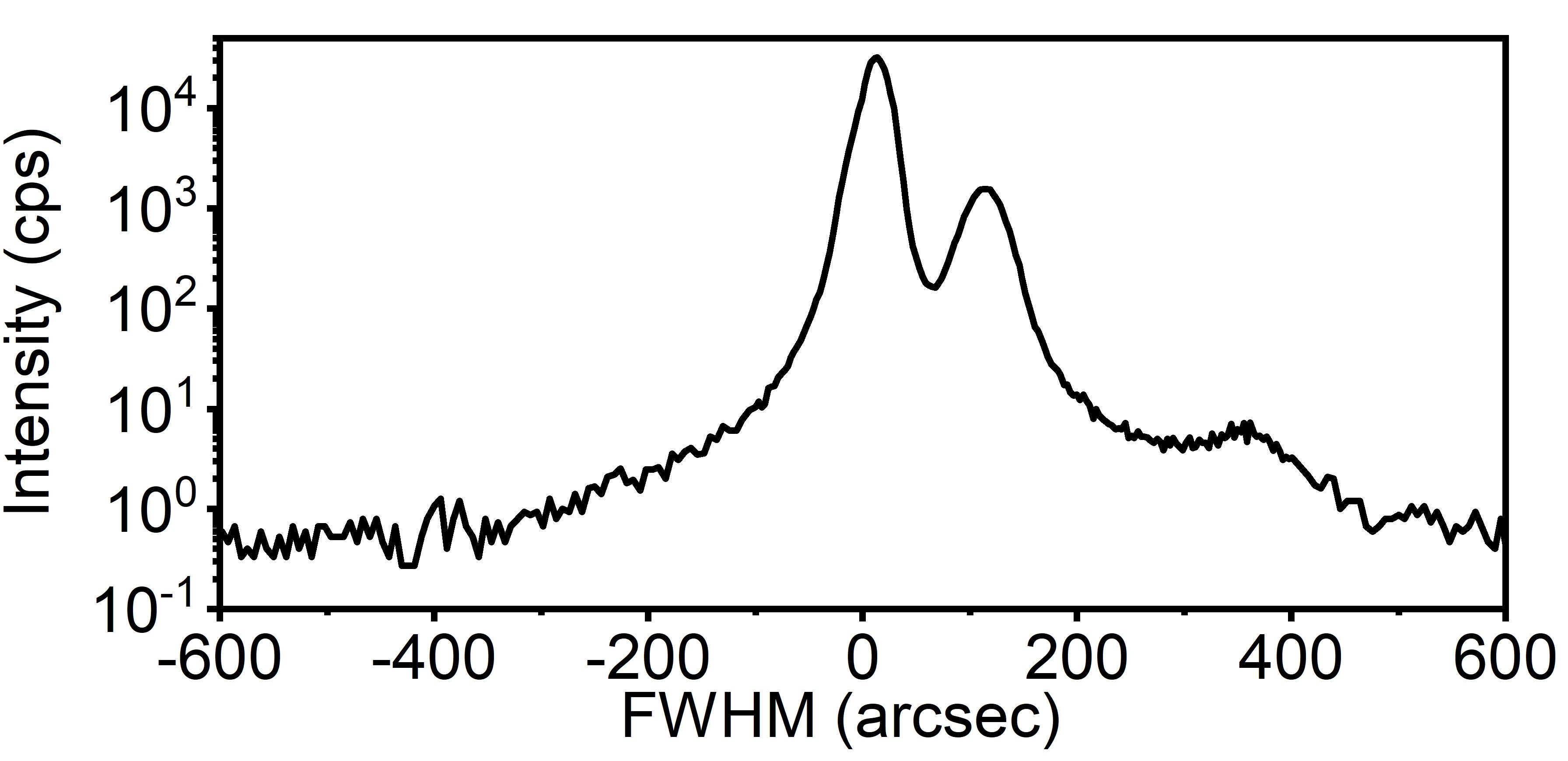}}
\caption{A triple-axis rocking curve measurement of the LN layer, validating wafer orientation and crystalline quality.}
\label{XRD}
\end{figure}

Scanning electron microscope images were obtained from a controlled die to validate the DRIE process, shown in Fig.~\ref{SEM} (a) and (b).  Fig.~\ref{SEM} (a) shows the Si cavity with the P3F LN structure, revealing a lateral over-etch at the cavity edge. The suspended LN layer in Fig.~\ref{SEM} (b) shows a smooth, continuous crystalline surface and a well-defined interface between the two LN layers. Moreover, the SEM confirms the 20 $\upmu$m LN layer thickness. The active layers were further examined using X-ray diffraction (XRD). Fig.~\ref{XRD_2theta} shows a double-axis $2\theta$–$\omega$ XRD scan of the active LN layer, confirming that the film is single phase and X-oriented $(11\bar{2}0)$, as only the $(11\bar{2}0)$ family of reflections is observed. The two LN layers superimposed in Fig.~\ref{XRD_2theta} can be separated using a triple-axis rocking curve measurement, shown in Fig.~\ref{XRD}. The rocking curve reveals a minor misorientation of the $(11\bar{2}0)$ surfaces with a value of 100$^{\prime\prime}$.  Furthermore, the rocking curve full-width-half-maxima (FWHM) are on the order of 40-50$^{\prime\prime}$, which is consistent with prior literature \cite{barrera_fundamental_2023, kramer_thin-film_2023, zheng_periodically_2025}.

\subsection{Measurements and Post-processing}

\begin{figure}[!t]
\centerline{\includegraphics[width=\columnwidth]{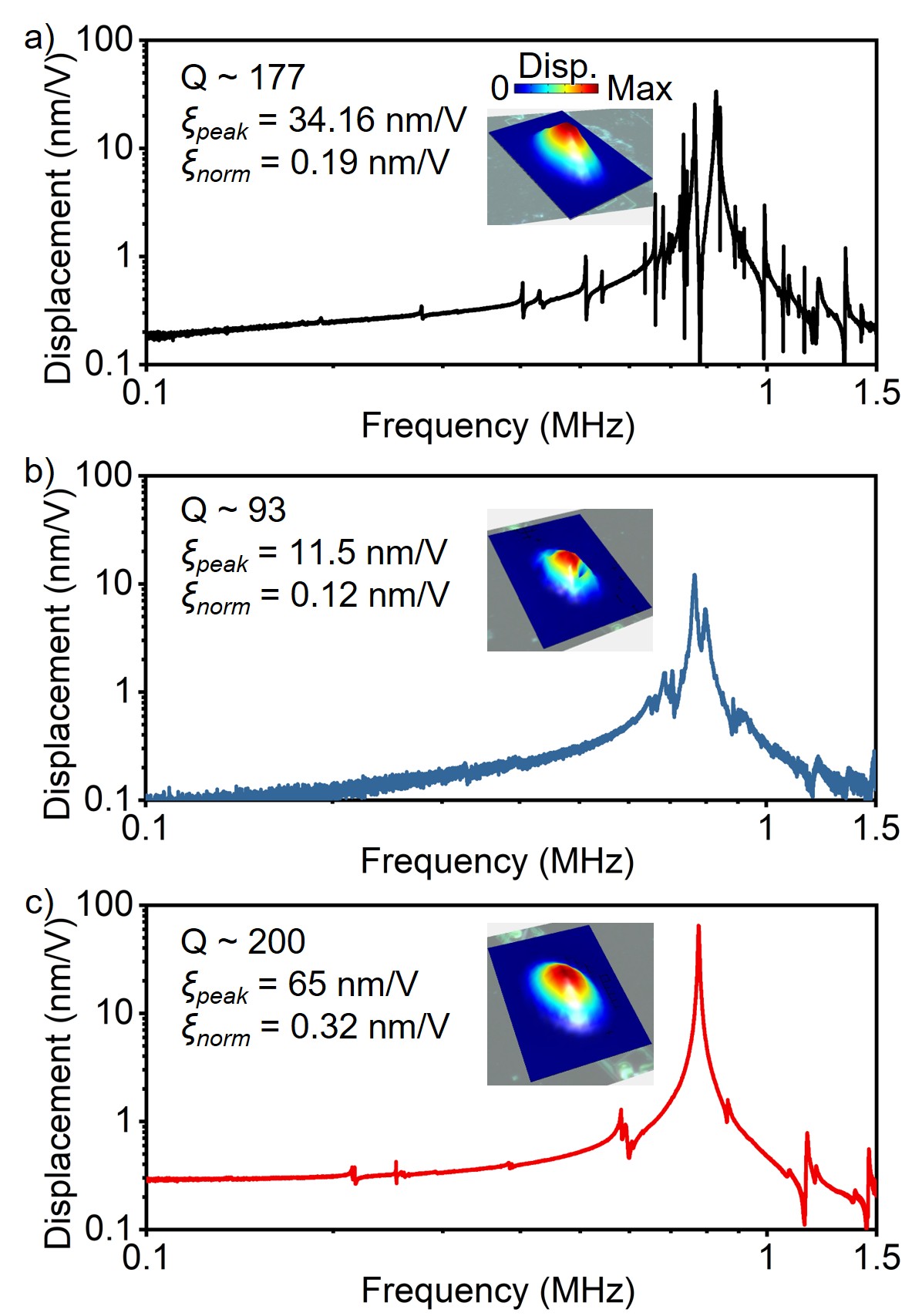}}
\caption{LDV measurements of the (a) unpackaged device, (b) packaged device, and (c) packaged device after thermal annealing. Insets show the vibrational mode shape and the extracted performance.}
\label{LDVMeas}
\end{figure}

The device was characterized both mechanically and electrically to assess transduction efficiency. Mechanical testing utilized a laser Doppler vibrometer (LDV) to evaluate the transducer's actuation capability. The electrical portion measured the device impedance and phase response using an impedance analyzer. These tests provided a comprehensive dataset to assess the \(Q\), \(k^2_\textit{eff}\), and any parasitic effects.

\begin{figure}[!t]
\centerline{\includegraphics[width=\columnwidth]{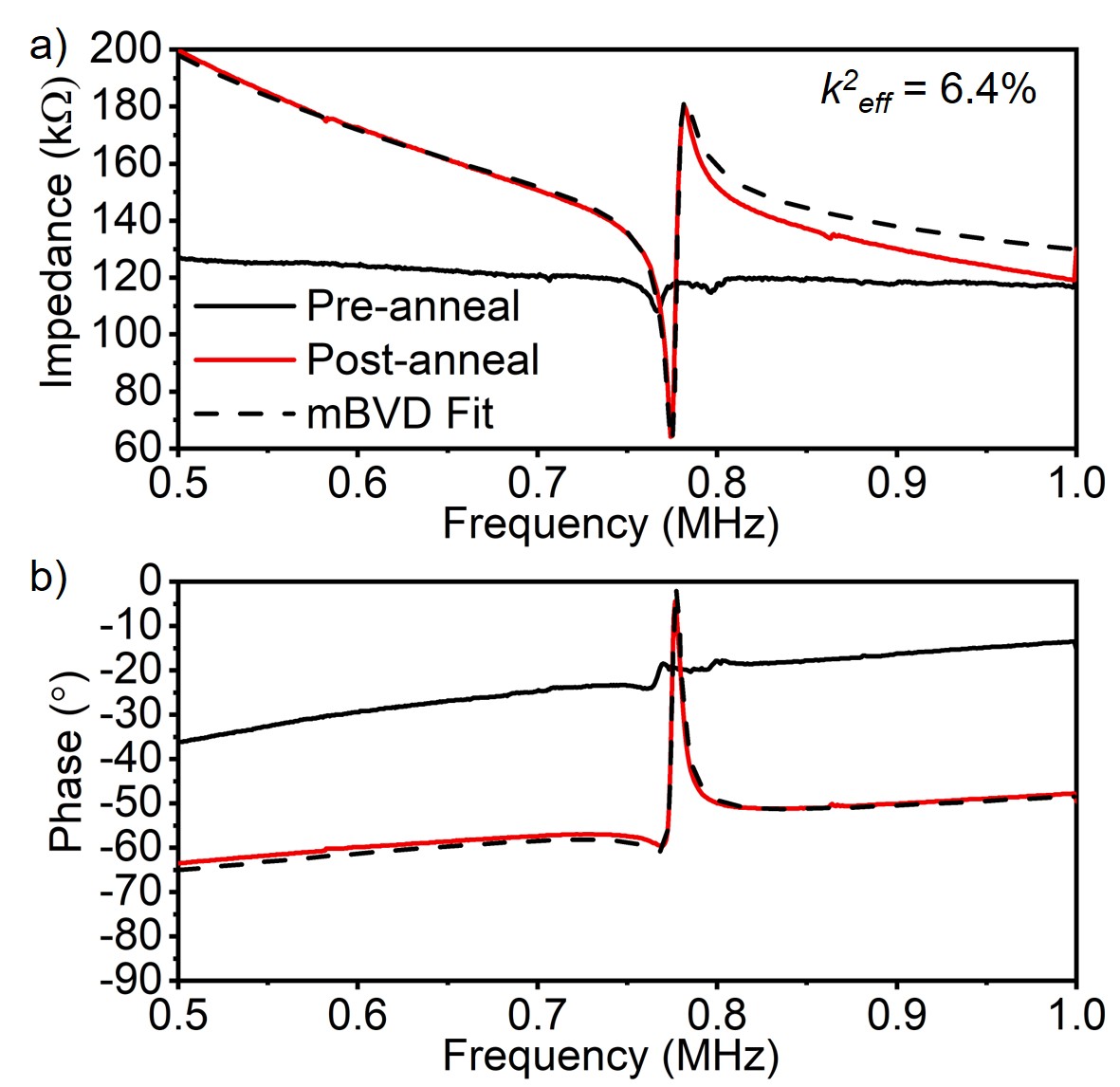}}
\caption{Electrical measurements of the LN PMUT depicting (a) impedance, and (b) phase. Measurements are performed before and after the post-processing steps.}
\label{ElecMeas}
\end{figure}

The mechanical characterization was performed using a Polytec PSV-500 sLDV (Scanning Laser Doppler Vibrometer) system. This system consisted of an integrated single-channel signal generator and a four-channel data acquisition module. The excitation signal applied to the device was a broadband chirp with a bandwidth of 5 MHz, and 5 averages were used to improve the signal-to-noise ratio (SNR). The excitation signal was amplified using a voltage amplifier (AALab A301HS). The fast Fourier transform (FFT) of the displacement waveform was normalized to the FFT of the input waveform to obtain the frequency response. The sLDV system automatically performed measurements at predefined points across the PMUT surface, and subsequent synthesis of all measurements provided the vibration profile at any selected frequency. The first mode shape is presented in Fig. \ref{LDVMeas} (a)–(c). The electrical measurements throughout the experiment were taken with a Zurich MFIA in a 2-terminal configuration, shown in Fig. \ref{ElecMeas} (a) and (b). 

\begin{figure}[!t]
\centerline{\includegraphics[width=0.8\columnwidth]{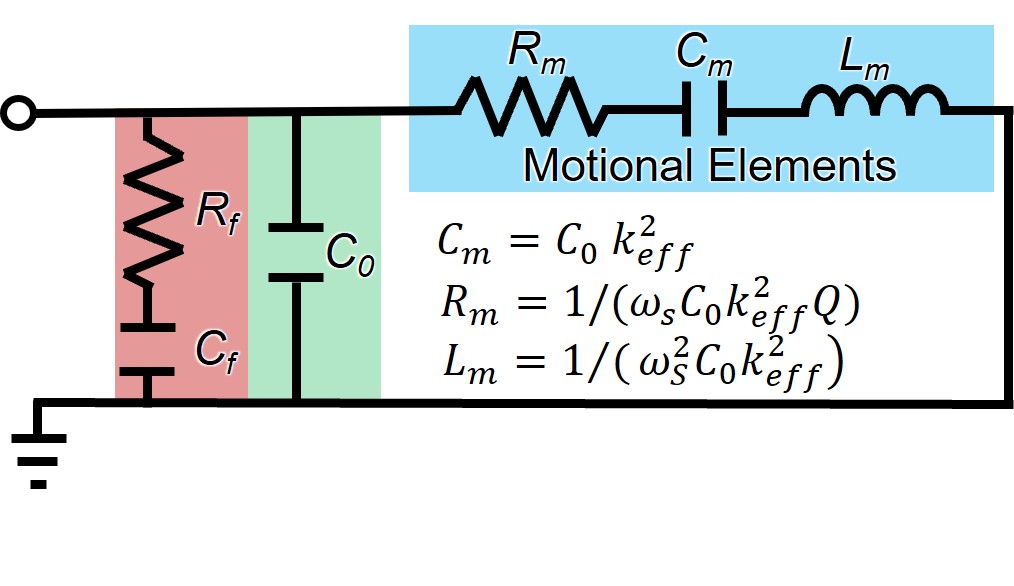}}
\caption{The mBVD model used to fit \(k^2_\textit{eff}\) in Fig. \ref{ElecMeas}. \(C_f\) and \(R_f\) are used to model parasitic feedthrough.}
\label{mBVD}
\end{figure}

Initial LDV measurements were conducted on an unpackaged die using direct current (DC) probes [Fig. \ref{LDVMeas} (a)]. The data confirms a flexural mode resonance but shows a high modal density, reduced displacement, and a lower operating frequency than predicted by the simulation (Fig. \ref{SimDisp}). To address these discrepancies, the die was subsequently mounted on a ceramic printed circuit board (PCB) using an epoxy adhesive. Electrical connections were made with Au wirebonds between the device electrodes and the Au PCB pads. The remeasured device [Fig. \ref{LDVMeas} (b)] shows fewer spurious modes alongside a further reduction in peak displacement.

Additionally, the impedance spectrum of the packaged device [Fig. \ref{ElecMeas} (a)] reveals a strong capacitive feedthrough, indicated by a nearly flat impedance profile over frequency. The phase response [Fig. \ref{ElecMeas} (b)] exhibits large dielectric loss, with the phase deviating significantly from the ideal 90$^\circ$. While the frequency shift may be attributed to the membrane size variation introduced by the DRIE process, the degraded performance and the cross-sectional images seen in Fig. \ref{SEM} (a) and (b) suggest intrinsic issues in the transducer stack. 

To mitigate performance degradation, the packaged device was thermally annealed in air using a Silicon Nitride (SiN) resistive heater. The temperature was ramped at approximately 25~$^\circ$C per minute to 400~$^\circ$C, held for 1 h, and then ramped back to room temperature. The anneal was performed in an open-air environment to avoid potential contamination from the organic epoxy used in the package. 

Post-anneal mechanical measurements [Fig. \ref{LDVMeas} (c)] show a well-defined primary resonance mode, with the peak displacement amplitude approaching its simulated value. The measured \(\xi_\textit{peak}\) was 65 nm/V at a \(Q\) of 200 at 775 kHz. Accordingly, the device yields an \(\xi_\textit{norm}\) of 0.325~nm/V. In comparison, the simulation predicted 0.354~nm/V at 1 MHz. Post‑measurement simulations ascribed the frequency shift to a ~34 $\upmu$m edge over‑etch per side. The improved mechanical performance is consistent with the post-anneal electrical measurements [Fig. \ref{ElecMeas} (a) and (b)], showing a significant improvement in the resonant device characteristics. However, although the parasitic feedthrough was largely mitigated, the phase response remained below 90$^\circ$. Nonetheless, while an additional anneal could potentially improve performance, no further post-processing steps were taken, and the post-anneal electrical data were analyzed to extract \(k^2_\textit{eff}\).

\begin{figure}[!t]
\centerline{\includegraphics[width=\columnwidth]{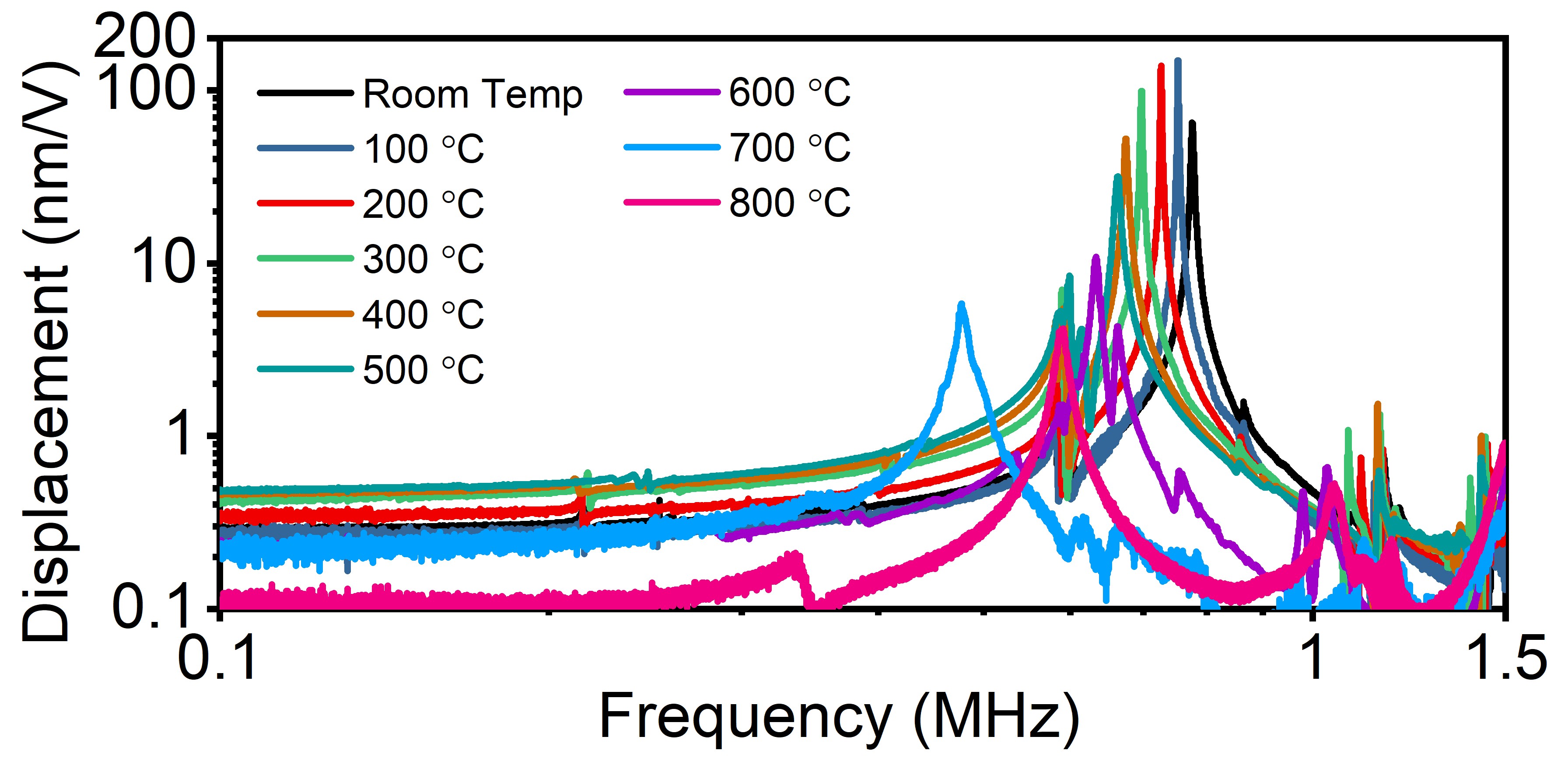}}
\caption{LDV measurements of the device from room temperature to 800~$^\circ$C.}
\label{Temperature}
\end{figure}

To evaluate electromechanical performance, the post-anneal electrical measurements were fit with a modified Butterworth-Van Dyke equivalent circuit model (mBVD) \cite{van_dyke_piezo-electric_1928}. The fit, shown as a dashed line in Fig. \ref{ElecMeas} (a) and (b), is obtained from the circuit depicted in Fig. \ref{mBVD}. A parasitic capacitor (\(C_f\)) in series with a resistor (\(R_f\)) is used to model the feedthrough network. The simulated device \(C_0\) was utilized to obtain their values, yielding a \(C_f\) of 1.63 pF and an \(R_f\) of 104 k$\Omega$. Although the model does not uniquely represent all parasitic effects, it fits the measured impedance and phase sufficiently well to extract the key performance metrics as follows:
\begin{equation}
k^2_\textit{eff} =  \frac{C_m}{C_0}
\label{eq:k2BVD}
\end{equation} 
This equation captures \(k^2_\textit{eff}\) from the mBVD model as the ratio of the mutual energy to the system's total energy. Hence, while their formulae differ, the two metrics from (\ref{eq:k2}) and (\ref{eq:k2BVD}) are equivalent \cite{beranek_acoustics_2019}. The extracted \(k^2_\textit{eff}\) was 6.4\%, showing a significant reduction from the simulated 11.8\%. Alongside the reduced operating frequency, the reduced \(k^2_\textit{eff}\) was attributed to a lateral overetch during the DRIE process, sufficiently altering \(D_1\) and \(D_2\). This discrepancy was further investigated and validated through post-measurement simulations, as discussed in the following section. 

Following measurement validation and post-processing, the LN PMUT demonstrated high performance. Despite a lower resonance frequency and smaller \(k^2_\textit{eff}\) than predicted, the device exhibited excellent efficiency. Given its mechanically robust geometry, the packaged device and the SiN heater were used to assess device performance at elevated temperatures. 

\subsection{Temperature Handling}

\begin{figure}[!t]
\centerline{\includegraphics[width=\columnwidth]{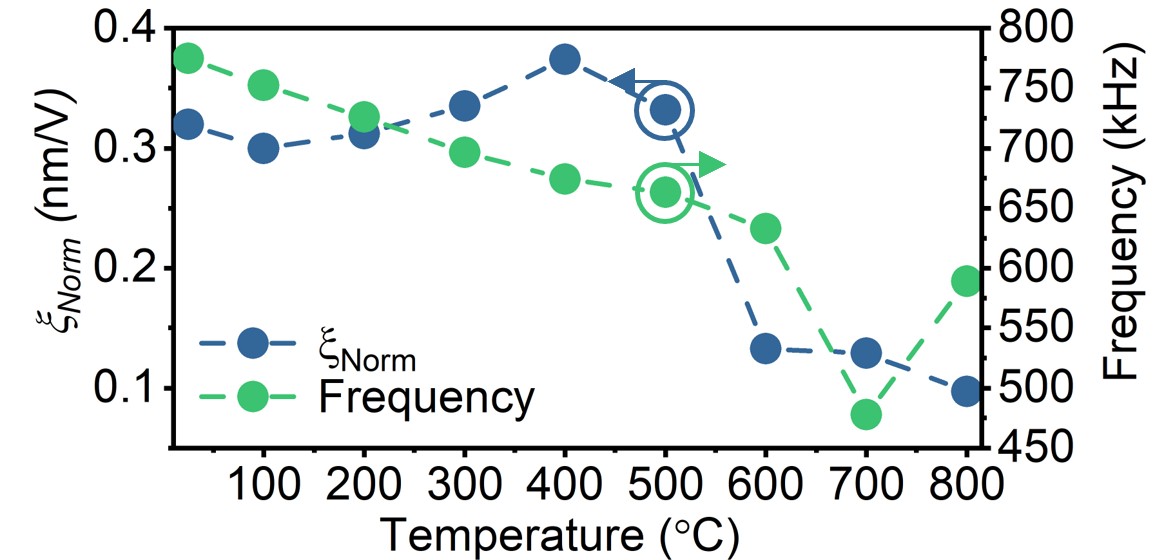}}
\caption{Extracted resonance frequency and \(\xi_\textit{norm}\) against temperature, highlighting stable operation with a high transduction efficiency up to 600~$^\circ$C.}
\label{TemperatureDoubleY}
\end{figure}

Thermal robustness was evaluated using the same SiN heater and LDV setup in open air. The temperature was ramped at 25~$^\circ$C per minute, and measurements were recorded every 100~$^\circ$C, until failure. These measurements, shown in Fig. \ref{Temperature}, demonstrate stable operation up to 600~$^\circ$C and structural survival up to 900~$^\circ$C. 

\begin{figure}[!t]
\centerline{\includegraphics[width=\columnwidth]{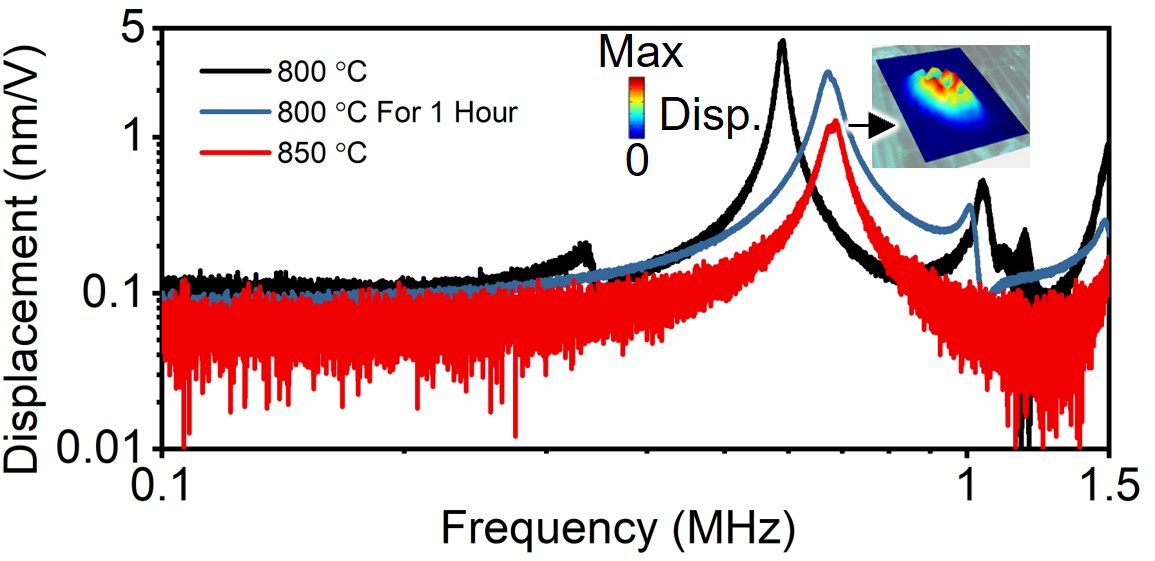}}
\caption{LDV measurements of the LN PMUT at 800~$^\circ$C and beyond, showing a slight frequency shift after sustained operation at 800~$^\circ$C for an hour.}
\label{800Cfor1hour}
\end{figure}
 \begin{figure}[!t]
\centerline{\includegraphics[width=\columnwidth]{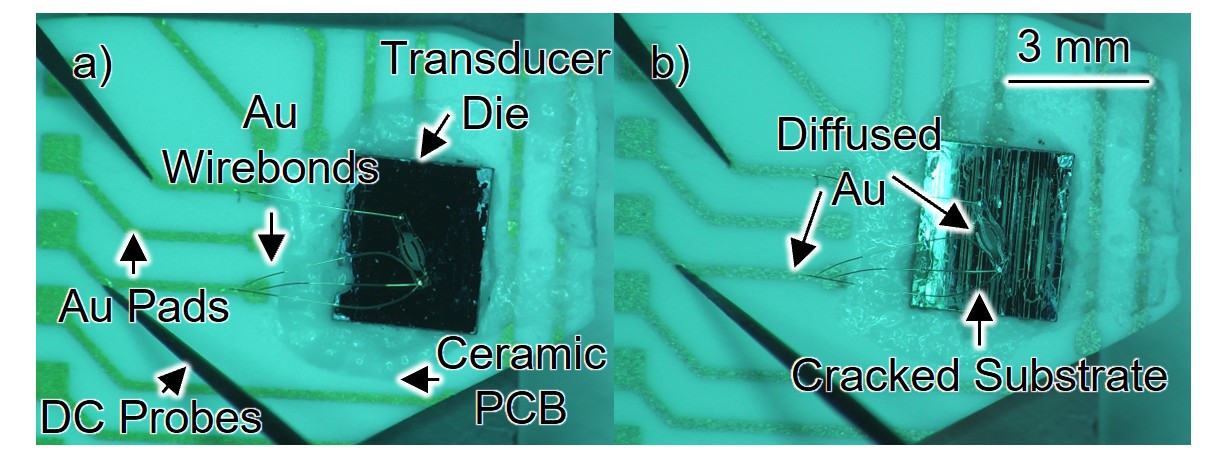}}
\caption{Optical images of the packaged LN transducer at (a) room temperature and (b) 900~$^\circ$C, depicting the experimental setup alongside the failure mechanism.}
\label{TemperatureOptical}
\end{figure}

 Mechanical measurements up to 600~$^\circ$C indicate a consistent frequency offset with increasing temperature, without loss in performance, as shown in Fig. \ref{TemperatureDoubleY}. At 500~$^\circ$C, the PMUT obtained a high \(\xi_\textit{norm}\) of 0.332 nm/V at 663~kHz, compared to the room-temperature value of 0.325 nm/V at 775~kHz. The extracted temperature coefficient of frequency (TCF) obtained by linear fitting over the range from room temperature to 600~$^\circ$C is -319 ppm/K. The number is considerably larger than the reported number near room temperature, but given a lack of complete material constant extraction for LN, the number here provides an experimental dataset \cite{li_temperature_2019}.
 
 Between 500~$^\circ$C and 600~$^\circ$C, the Si substrate exhibited cracking, attributed to thermal stress. However, the active layer showed no visible damage and continued to operate in flexural mode. Substrate failure resulted in an anomalous frequency shift from 633~kHz at 600~$^\circ$C to 476~kHz at 700~$^\circ$C. With continued cracking of the Si substrate and epoxy reflow, the membrane boundary conditions were recovered between 700~$^\circ$C and 800~$^\circ$C, showing 4.2 nm/V at 590~kHz with a \(Q\) of 43, leading to an \(\xi_\textit{norm}\) of 0.097 nm/V. The device sustained 800~$^\circ$C for an hour (Fig.~\ref{800Cfor1hour}), and failed at temperatures approaching 900~$^\circ$C as the electrodes began to diffuse into the environment. The final recorded measurement at 850~$^\circ$C shows further recovery of the resonance frequency and a flexural mode shape (Fig.~\ref{800Cfor1hour}). The optical images of the device and the measurement setup at room temperature and at 900 $^\circ$C can be seen in Fig. \ref{TemperatureOptical} (a) and (b), respectively. The optical images at 900 $^\circ$C show substrate failure around the diffused gold electrodes and pads. Continued device survival at high temperatures and the associated failure mechanisms will be closely studied in the future.

\section{Discussion}

The bimorph LN PMUT reveals both strengths and remaining limitations of the P3F LN platform for ultrasonic transduction. High performance and temperature resilience were achieved through extensive design considerations and post-processing steps; however, several performance and fabrication detriments remain. Understanding these effects, ranging from the frequency offset to \(k^2_\textit{eff}\) loss and remaining feedthrough, is essential for further platform development. This section addresses the measured frequency shift via post-measurement simulations and analyzes its impact on \(k^2_\textit{eff}\) to obtain a validated model. The updated simulation is then used to gain further insights into the bimorph LN PMUT platform by predicting receiving sensitivity. The complete dataset better justifies the design choices and elaborates on the obtained performance.   

\subsection{Post-measurement Simulations}

\begin{figure}[!t]
\centerline{\includegraphics[width=\columnwidth]{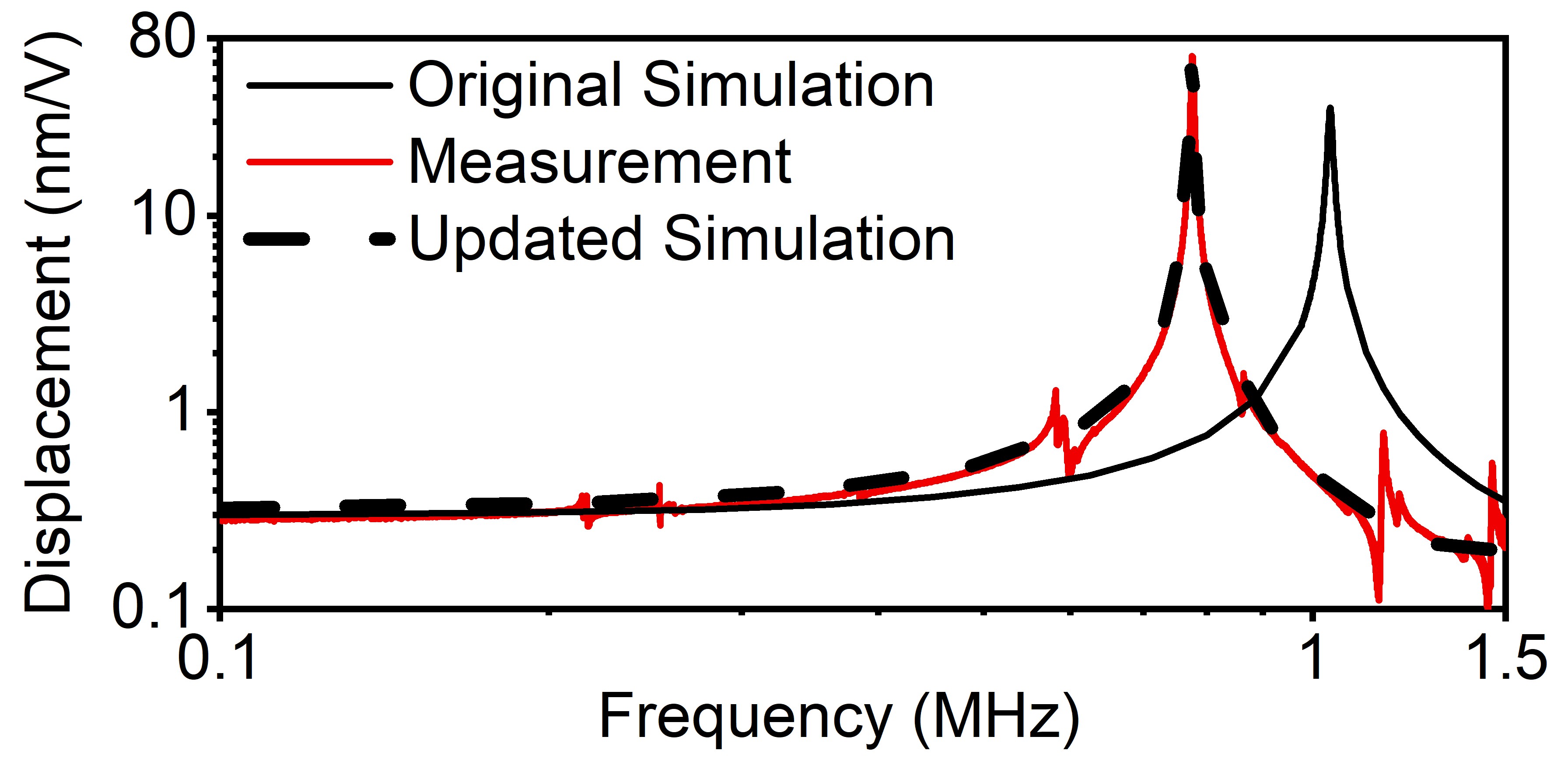}}
\caption{Superimposed plots of the measured displacement, original simulated displacement, and its updated version.}
\label{UpdatedSim}
\end{figure}

The variation in the measured device frequency against the simulation was attributed to a lateral overetch from the DRIE process, as supported by Fig. \ref{fab} (b) and Fig. \ref{SEM} (a). Incorporating a 33.8 $\upmu$m lateral overetch on each side of the membrane brings the simulated resonance frequency in agreement with its measured value. In Fig.~\ref{UpdatedSim}, the updated simulation yields an \(\xi_\textit{norm}\) of 0.345 nm/V at 775 kHz, which closely matches the measured value of 0.325 nm/V at the same resonance frequency. 

The lateral overetch has a pronounced effect on \(k^2_\textit{eff}\). The simulated value is reduced from 11.8\% to 6.5\%. Due to the minimal impact of the overetch on \(C_0\), the extracted \(k^2_\textit{eff}\) of 6.4\% remains, in good agreement with the simulation. The remaining minor discrepancy likely stems from the limitations of the mBVD fit. In particular, the lossy feedthrough branch, modeled as a simple RC series network, cannot fully capture both impedance and phase across the entire frequency band [Fig. \ref{ElecMeas} (a) and (b)]. Although additional parasitic elements may improve the fit, their physical significance would be ambiguous. 

\begin{figure}[!t]
\centerline{\includegraphics[width=\columnwidth]{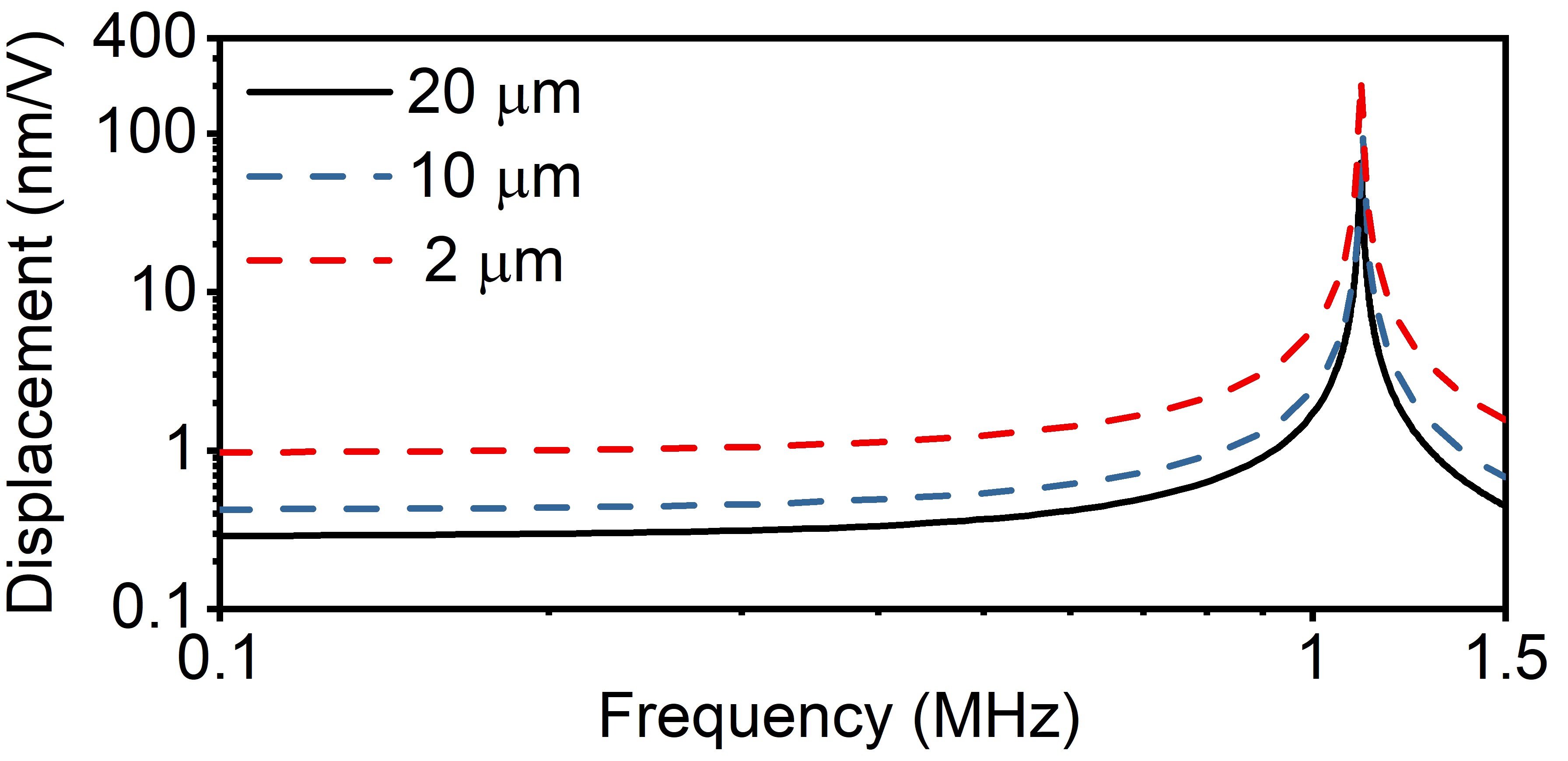}}
\caption{Changing transmit efficiency at 1 MHz with a changing piezoelectric film thickness, highlighting its enhancement with a thinner active layer. The simulations utilize a \(Q\) of 200.}
\label{VarThicknessDisp}
\end{figure}

Post-measurement simulations were utilized to evaluate the dependence of PMUT transmit efficiency on the active layer thickness, which remains its primary limiting factor in the present geometry. Fig. \ref{VarThicknessDisp} shows the optimized design scaled linearly with respect to the piezoelectric film thickness, illustrating varying performance at 1 MHz. These results indicate that transmit efficiency in LN PMUTs can be significantly enhanced. For instance, a 2 $\upmu$m bimorph active layer enhances \(\xi_\textit{norm}\) from 0.354 nm/V to 1.14 nm/V. Given the close agreement between simulation and measurement, Fig. \ref{VarThicknessDisp} enables a meaningful comparison with the SoA. For reference, a 3 $\upmu$m thick air-coupled PZT PMUT can obtain a \(\xi_\textit{norm}\) of 4.87 nm/V at 1 MHz \cite{xing_design_2020}. As a result, although high-\(d\) materials, such as PZT, outperform LN in transmit efficiency at comparable compliance levels, LN provides sufficient actuation capability  for high sound pressure level (SPL) operation while featuring high receive sensitivity at the same time.

\subsection{Receiving Performance}

\begin{figure}[!t]
\centerline{\includegraphics[width=\columnwidth]{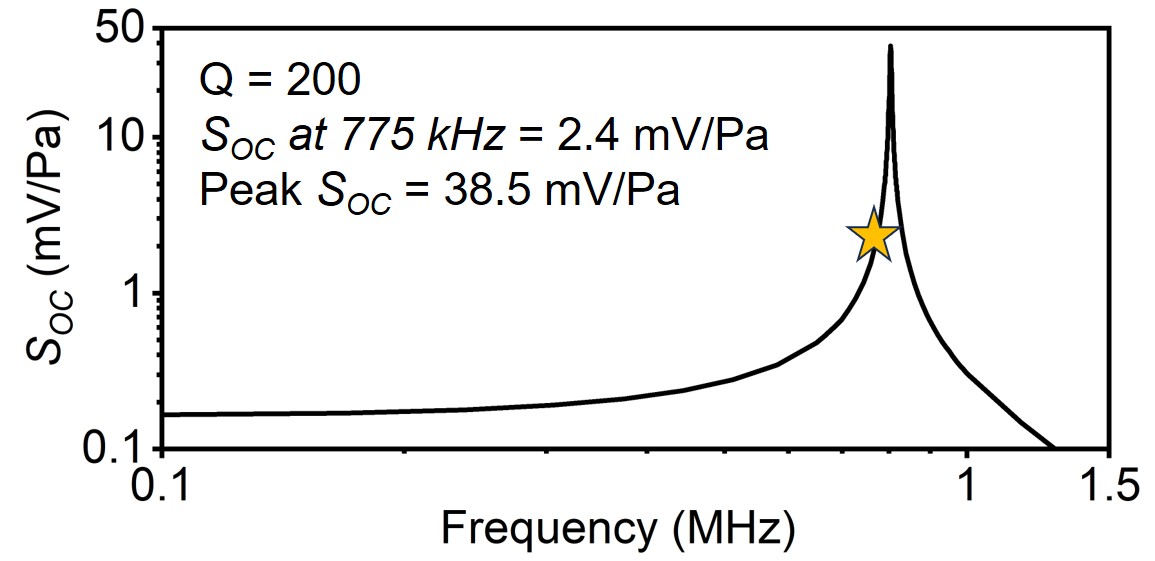}}
\caption{\(S_\textit{OC}\) obtained using the post-measurement simulation, showing a close agreement with the value calculated using measured data, indicated by a star.}
\label{VOCSim}
\end{figure}

Given the strong agreement between measurements and the updated simulation, the FEA can be used directly to extrapolate the receiving performance of the bimorph LN PMUT. Using reciprocity in piezoelectric systems, the open-circuit voltage sensitivity (\(S_\textit{OC}\), in V/Pa) can be obtained with the ratio between the simulated specific acoustic compliance (\(C_\textit{s}\), in m/Pa) and the measured \(\xi_\textit{peak}\) \cite{hunt_electroacoustics_1982}. Hence, the open-circuit voltage can be found as below:
\begin{equation}
S_\textit{OC} =  \frac{C_s}{\xi_\textit{peak}}  
\label{eq:OCV}
\end{equation} 
For the measured 65 nm/V peak actuation displacement at a \(Q\) of 200, a \(C_s\) of 160 pm/Pa is simulated. Accordingly, the resulting open-circuit sensitivity (\(S_\textit{OC}\)) is 2.46 mV/Pa at 775 kHz. To validate the calculated value, we compare it to the open-circuit voltage obtained directly from the FEA, shown in Fig. \ref{VOCSim}. The simulation confirms the calculated estimate, showing 2.4 mV/Pa at 775 kHz. Additionally, Fig. \ref{VOCSim} shows that the device achieves a peak open-circuit voltage at its parallel resonance, reaching up to 38 mV/Pa. As in the case of transmit efficiency, we decouple sensitivity from \(Q\) by utilizing the normalized open-circuit sensitivity (\(S_\textit{OC}^\textit{norm}\)). Hence, we obtain a \(S_\textit{OC}^\textit{norm}\) of 12 $\upmu$V/Pa. Nonetheless, sensitivity is highly influenced by the interfacing circuitry, and its effect is most pronounced at the system level. However, these results cannot be experimentally validated due to the electrode failure seen in Fig. \ref{TemperatureOptical}, motivating further studies of bimorph LN-based sensors.

While external factors play a large role in receiving performance, a comparison with incumbent PMUT platforms highlights the LN PMUT's high voltage sensitivity. State-of-the-art 36\% ScAlN-based air-coupled PMUT shows an \(S_\textit{OC}\) of 1 mV/Pa at 60 kHz via a pulse-echo measurement \cite{kusano_high-spl_2019}. On the other hand, its air-coupled PZT counterpart for long-range detection shows 2 mV/Pa at 40 kHz through a pitch-catch experiment \cite{luo_airborne_2021}. The performance of these ferroelectric platforms is further enhanced by incorporating a DC bias. For example, the sensitivity in a 15\% ScAlN PMUT is increased from 6.43 mV/Pa in air at 150 kHz to 9.67 mV/Pa with an 80 V bias \cite{choong_dc_2023}. As an alternative, PMUTs can also utilize a dual-electrode structure to improve receiving performance. For instance, \cite{liu_dual-electrode_2024} shows 4.4 mV/Pa at 55 kHz in 30\% ScAlN. By comparison, the bimorph LFE LN PMUT achieves a comparable \(S_\textit{OC}\) while operating at a higher frequency, showing potential for a highly sensitive, robust PMUT platform.

\subsection{Remaining Limitations and Future Development}

The LFE LN PMUT relied on post-processing steps, such as packaging and annealing, to achieve high performance. Understanding the underlying issues that necessitate these steps is valuable for future LN PMUT implementations. A closer look at the spurious modes excited in the unpackaged transducer [Fig. \ref{LDVMeas} (a)] reveals that the unwanted tones were also flexural modes, actuating different parts of the substrate. Therefore, the high modal density seen in Fig. \ref{LDVMeas} was attributed to weak energy confinement between the stiff 20~$\upmu$m LN and the relatively softer 200~$\upmu$m thick Si. Consequently, the packaging constrained the substrate, suppressing spurious modes [Fig. \ref{LDVMeas} (b)]. Despite this improvement, the packaged device still exhibited poor mechanical and electrical performance, which was mitigated by a 400~$^\circ$C anneal in open air. While further studies are necessary to determine the exact mechanism, the improvement is preliminarily attributed to stress relaxation and possible changes in contact and interfaces; the exact mechanism remains to be investigated. 

The remaining parasitic feedthrough observed in the post-anneal electrical measurements (Fig. \ref{ElecMeas}) is critical for future LN sensor development. Although LN intrinsically exhibits low dielectric loss, parasitic feedthrough acts as a major bottleneck, limiting achievable performance. While this effect was partially mitigated by annealing, residual parasitics constrain the usable sensitivity, particularly near the parallel resonance. While thermal treatment appears beneficial, the fundamental origin of the feedthrough remains unresolved and warrants further studies.

Beyond addressing parasitic effects, the thickness scalability of the P3F LN platform presents additional opportunities for optimization. The present structure features a thick P3F piezoelectric layer to prioritize robustness and voltage sensitivity, however, Fig. \ref{VarThicknessDisp} shows that reducing film thickness can substantially enhance transmit efficiency by increasing membrane compliance. P3F LN supports this flexibility, with high-quality P3F films demonstrated at thicknesses ranging from 100 nm to 300 $\upmu$m \cite{barrera_50_2026, cho_238-ghz_2024, kramer_acoustic_2025, matto_direct_2023}. As a result, future P3F LN PMUT implementations will focus on thickness scaling to maximize transmit efficiency while maintaining sufficient mechanical integrity for required operating conditions.

\section{Conclusion}
This work introduces bimorph LFE LN PMUTs for ultrasonic transduction, demonstrating high performance alongside extreme temperature resilience. Motivated by the comparison drawn between LN and incumbent PMUT material platforms, the study leverages P3F LN to design bimorph PMUTs. A 775 kHz flexural mode device was fabricated and measured with a 20 $\upmu$m thick active layer, yielding a high transmit efficiency with \(\xi_\textit{norm}\) of 0.325 nm/V and a high voltage sensitivity of 2.4 mV/Pa. The device showed mechanical and thermal robustness, maintaining high performance up to 600~$^\circ$C and survival to 900~$^\circ$C. The analysis performed throughout the experiment provides insights into the LFE LN PMUT platform and isolates future study steps. With these promising results, the proposed material platform shows significant promise for rugged sensor applications as well as bidirectional operation. 

\section*{Acknowledgment}
The authors would like to thank Dr. Todd Bauer and Dr. Kwok-Keung Law for helpful discussions. The authors would also like to thank Dr. Bichoy Bahr, Dr. Udit Rawat, and Dr. Yao Yu from the Texas Instruments Kilby Labs for valuable feedback. 

\bibliographystyle{IEEEtran}
\bibliography{thebibliography}

\end{document}